\documentclass[journal=accacs,manuscript=article]{achemso}
\usepackage[version=3]{mhchem}
\usepackage{graphicx}
\usepackage{bm}
\usepackage[T1]{fontenc}
\usepackage{sourceserifpro,sourcecodepro,sourcesanspro}
\usepackage[utf8]{inputenc}
\usepackage[protrusion=true,expansion=true,final]{microtype}
\usepackage{mathastext}
\usepackage{booktabs}
\usepackage{mciteplus}
\mciteErrorOnUnknownfalse
\usepackage[usenames,dvipsnames]{xcolor}
\usepackage{xr}
\usepackage{filecontents}
\usepackage[bookmarks=false,colorlinks]{hyperref}
\hypersetup{
	linkcolor= DeepRed,        
	citecolor=DeepRed,        
	filecolor=DeepRed,        
	urlcolor=DeepRed,         
	pdfauthor={Debangshu Mukherjee, Jocelyn T.L.  Gamler, Sara E. Skrabalak, Raymond R. Unocic},
	pdftitle={Lattice strain measurement of core@shell electrocatalysts with 4D-STEM nanobeam electron diffraction},
}
\usepackage{xfrac}

\definecolor{DeepRed}{rgb}{0.5,0,0}

\author{Debangshu Mukherjee}
\email{mukherjeed@ornl.gov}
\affiliation{Center for Nanophase Materials Sciences, Oak Ridge National Laboratory, Oak Ridge, TN, USA}

\author{Jocelyn T.L. Gamler}
\affiliation{Department of Chemistry, Indiana University -- Bloomington, Bloomington, IN, USA}

\author{Sara E. Skrabalak}
\affiliation{Department of Chemistry, Indiana University -- Bloomington, Bloomington, IN, USA}

\author{Raymond R. Unocic}
\email{unocicrr@ornl.gov}
\affiliation{Center for Nanophase Materials Sciences, Oak Ridge National Laboratory, Oak Ridge, TN, USA}

\title{Lattice strain measurement of core@shell electrocatalysts with 4D-STEM nanobeam electron diffraction}

\keywords{4D-STEM, electron microscopy, strain measurement, electrocatalysts, nanocatalysts}

\begin{document}

	\begin{abstract}
		Strain engineering enables the direct modification of the atomic bonding and is currently an active area of research aimed at improving the electrocatalytic activity. However, directly measuring the lattice strain of individual catalyst nanoparticles is challenging, especially at the scale of a single unit cell. Here, we quantitatively map the strain present in rhodium@platinum (core@shell) nanocube electrocatalysts using conventional aberration-corrected scanning transmission electron microscopy (STEM) and the recently developed technique of 4D-STEM nanobeam electron diffraction. We demonstrate that 4D-STEM combined with data pre-conditioning allows for quantitative lattice strain mapping with sub-picometer precision without the influence of scan distortions. When combined with multivariate curve resolution, 4D-STEM allows us to distinguish the nanocube core from the shell and to quantify the unit cell size as a function of distance from the core-shell interface. Our results demonstrate that 4D-STEM has significant precision and accuracy advantages in strain metrology of catalyst materials compared to aberration-corrected STEM imaging and is beneficial for extracting  information about the evolution of strain in catalyst nanoparticles.
	\end{abstract}
    \section{\label{sec:intro}Introduction}
    	The implementation of a clean, sustainable and low carbon future requires the development of high-performance electrocatalysts\cite{electrocatalyst_review}. Electrocatalysts are used to catalyze electrochemical reactions, which are predominantly redox reactions that occur on an electrode surface\cite{electrocat_book}. Examples of such reactions include the cathodic oxygen reduction reaction (ORR) and the hydrogen evolution reaction (HER)\cite{ORR_electrocat,HER_electrocat}. Most electrocatalysts are synthesized from noble metals such as platinum, rhodium or palladium\cite{electrocat}. In contrast to widely available metals such as iron (56,300 ppm) or titanium (6,200 ppm), the concentrations of noble metals in the earth's crust is several orders of magnitude lower -- approximately 0.001 pm for rhodium, 0.005 ppm for platinum and 0.015 ppm for palladium, making them significantly more expensive\cite{abundance_book}. Therefore, as a result of the cost and abundance issues, a primary goal of noble metal catalysis research has been to develop strategies to increase the catalyst \textit{mass activity}, i.e., the catalytic activity per unit mass. Since catalysis is an inherently surface driven phenomenon, research has been focused on two complementary goals; increasing the surface to volume ratio and increasing the catalytic activity of the nanoparticle surfaces. Since decreasing the particle size increases the surface to volume ratio, with the absolute limit being reached by a single atom, research in this area has been focused on the synthesis of catalyst nanoparticles with different morphologies such as nanocubes, nanocages, etc. rather than bulk catalysts\cite{Au_nano_comparison,platinum_nanocages,nanocage2,nanobox,nanobelt,concave_nanocube}. 
    	
    	Toward the goal of increasing surface catalytic activity, the strength of surface-adsorbate interactions and, in turn reaction rates, can be tuned through changes in the electronic structure of the metal surface, which can be achieved by modification of the surface strain where the degree of metal orbital overlap is tunable\cite{strain_review,theoretical_strain_core_shell,strain_tuning}. The impact of surface strain on catalysis has been explained by the \textit{d-band} center model, where an increase of the inter-atomic distances (i.e., metal orbital overlap) shifts the \textit{d-band} center to a lower energy as compared to the unstrained metal and weakens the surface-adsorbate interactions\cite{d_band}. In contrast, a decrease in the inter-atomic distances shifts the \textit{d-band} center to a higher energy and consequently strengthens the surface-adsorbate interactions. Multiple synthesis methods have been proposed to increase catalytic activity on the surface through strain engineering – such as argon bombardment of the surface \cite{strain_surface_tio2}, alloyed and/or intermetallic nanoparticles\cite{jocelyn_sara_progress,intermetallic,alloy,Pt_strain_ORR,intermetallic_np_order}, or by the epitaxial deposition of a metallic shell to create core@shell nanoparticles where the lattice mismatch between the particle core and the shell creates a strained outer layer\cite{core_shell_SERS,core_shell_AuPd,core_shell_CoPt,lattice_strain,lattice_strain_focal_series,insitu_NP_strain,skrabalak_bimetallic_review,alloyed_shell}.
    	
    	In principle, the strain in the nanoparticle shell is directly related to the lattice mismatch between the core and shell; thus, core@shell nanoparticles are ideal platforms for tuning surface catalysis, where the degree and nature of the lattice strain (compressive vs. tensile) can be modified via the extent of lattice mismatch and shell thickness. Recent theoretical and experimental investigations, however, have shown that strain engineering in such systems is significantly more complicated than encompassed by a simplistic picture of lattice mismatch\cite{core_shell_stress_release}. Three-dimensional (3D) measurements of lattice strain performed by reconstructing atom positions from annular dark field (ADF) scanning transmission electron microscopy (STEM) or by combining electron tomography and ADF-STEM imaging have demonstrated that strain states are established by the distance from the surface, surface faceting, local chemical environment, presence of nanocrystalline grains, and so on\cite{lattice_strain_3D,nanoparticle_tomography}. This was also shown through theoretical modeling where large-scale molecular dynamics (MD) simulations of Pd@Au nanoparticles exhibited compressive stresses within the Pd core and strain states in the Au shell arose from a complex interplay between the interface distance, distance from the particle surface, and the crystallographic orientation\cite{CDI}.
    	
    	Reported measurement of strain distributions in nanoparticles are sparse and are experimentally challenging. The most straightforward way to measure strain in bulk materials is by X-ray diffraction (XRD), where the lattice parameter can be calculated from the diffraction peaks and strain is measured by comparing the calculated experimental parameters to known values. However, conventional XRD runs into resolution problems for nanoparticles due to particle size effects, with recent coherent diffraction imaging strain mapping experiments achieving a  resolution of a few nanometers\cite{CDI}, which is on the order of the size of the nanoparticles themselves, making it impossible to distinguish strain in the shell from the particle core. Yet another issue with using XRD methods to measure strain in nanoparticles is the extreme monodispersity required for the sample size and shape, which is hard to achieve when synthesizing complex nanoparticles\cite{nanocube_pd_xrd}.
    	
    	In this work, strain engineered bimetallic Rh@Pt nanocubes, which have been previously found to be excellent electrocatalysts for formic acid oxidation was used as a model system. In this system, it was determined that modification of the Rh shell thickness results in a variation of compressive surface strain and alters the electronic structure\cite{skrabalak_RhPt,josie_sara}. A combination of two electron microscopy techniques: aberration-corrected STEM imaging and 4D-STEM nanobeam electron diffraction were used to quantitatively measure strain distributions. We compare the results from both techniques and demonstrate how 4D-STEM can offer a superior strain metrology approach across the core@shell interface.
    
    \section{\label{sec:mat_meth}Materials and Methods}
      \subsection{\label{ssec:sample_prep}Preparation of core@shell Rh@Pt nanocubes}
      
      \textbf{Chemicals:} Polyvinylpyrrolidone (55,000 M.W., PVP) and platinum (II) acetylacetonate $\mathrm{\left(Pt\left(acac\right)_2\right)}$ were purchased from Aldrich. Rhodium(III) bromide hydrate $\mathrm{\left(RhBr_3.xH_2O\right)}$, and triethylene glycol (TREG), was purchased from Alfa-Aesar. Ethylene glycol (anhydrous, 99.8\%, EG) purchased from Sigma-Aldrich. All chemicals were used without further purification.
      
      \textbf{Rh nanocubes:} Synthesis of Rh nanocubes was adapted from a report by Biacchi \textit{et. al.}\cite{rh_nanoparticles}. 102 mg $\mathrm{\left(RhBr_3.xH_2O\right)}$ was placed in a vial with enough ethanol to completely dissolve the rhodium salt. The solution was then placed in a 50 mL three-necked round bottom flask with 230 mg PVP (55,000 MW) and 10.0 mL of TREG. Argon gas was continuously purged through the solution, and the reaction vessel was equipped with stir bar and a condenser. The solution temperature was heated to $\mathrm{110^{\circ}C}$ in an oil bath for 15 minutes to initiate nucleation. The temperature was then raised to $\mathrm{145^{\circ}C}$ for 90 minutes. The solution was allowed to cool to room temperature. The product was then washed with acetone and collected by centrifugation as previously described and re-dispersed in 10 mL ethanol.
      
      \textbf{Rh@Pt nanocubes:} Rh@Pt nanocubes were synthesized as reported by Harak \textit{et. al.}\cite{skrabalak_RhPt}. 1.0 mL of Rh cubic seeds and 10.0 mL of ethylene glycol was placed in a 50 mL three-necked round bottom flask. The reaction flask was equipped with a stir bar and a condenser to prevent any evaporation of the ethylene glycol. The solution was purged with argon gas as it was rapidly heated to $\mathrm{160^{\circ}C}$ over the course of 6-8 minutes. Meanwhile, the desired amount of $\mathrm{Pt\left(acac\right)_2}$ (5 mg for thinner Pt shell or 12 mg for thicker Pt shell) was placed in a vial and acetone was added until the salt had completely dissolved. Once the Rh cube/ethylene glycol solution had reached $\mathrm{160^{\circ}C}$, the $\mathrm{Pt\left(acac\right)_2}$ solution was rapidly hot-injected into the flask with a syringe, and the reaction was heated for two hours. The solution was allowed to cool to room temperature. The product was then washed with acetone, collected by centrifugation, and redispersed in 10 mL of ethanol.
    \subsection{\label{ssec:TEM}STEM of core@shell nanocubes} 
    The Rh@Pt nanocubes were sonicated and then deposited on amorphous carbon grids, followed by drying at room temperature. STEM characterization of the samples was performed using a Nion UltraSTEM 100 operated at an accelerating voltage of 100 kV. Aberration-corrected ADF-STEM images were acquired with a probe-forming convergence angle of 32 mrad and the images were collected using an ADF detector with collection angles from 84-200 mrad. Images were collected with a pixel dwell time of 4 $\mathrm{\mu}$s and a pixel spacing of 7.8 pm. To correct for scan distortions, a pair of STEM images was collected with two perpendicular fast scan directions, which were used to subsequently correct for scan drift using a procedure developed by Ophus \textit{et. al.}\cite{drift_corr}. For 4D-STEM experiments, the probe-forming convergence angle  was decreased from 32 mrad to approximately 5 mrad while keeping all the other microscope parameters constant. The nanodiffraction patterns were collected at every scan position using a Hamamatsu Orca CMOS detector with a pixel dwell time of 2 ms, which is 500 times slower than the ADF-STEM pixel dwell time. Since the electron beam size is proportional to the inverse of the condenser aperture angle, a coarser pixel sampling of 1 \r{A} was used for 4D-STEM nanodiffraction.
    
    The ADF-STEM and the 4D-STEM datasets were analyzed using custom-developed Python codes that are open sourced at Github\cite{stemtools}.
    \section{\label{sec:R_D}Results and Discussions}
    \subsection{\label{ssec:adf_stem}Measuring strain with atomic resolution STEM} 
    	\begin{figure*}
	    	\includegraphics[width=\textwidth]{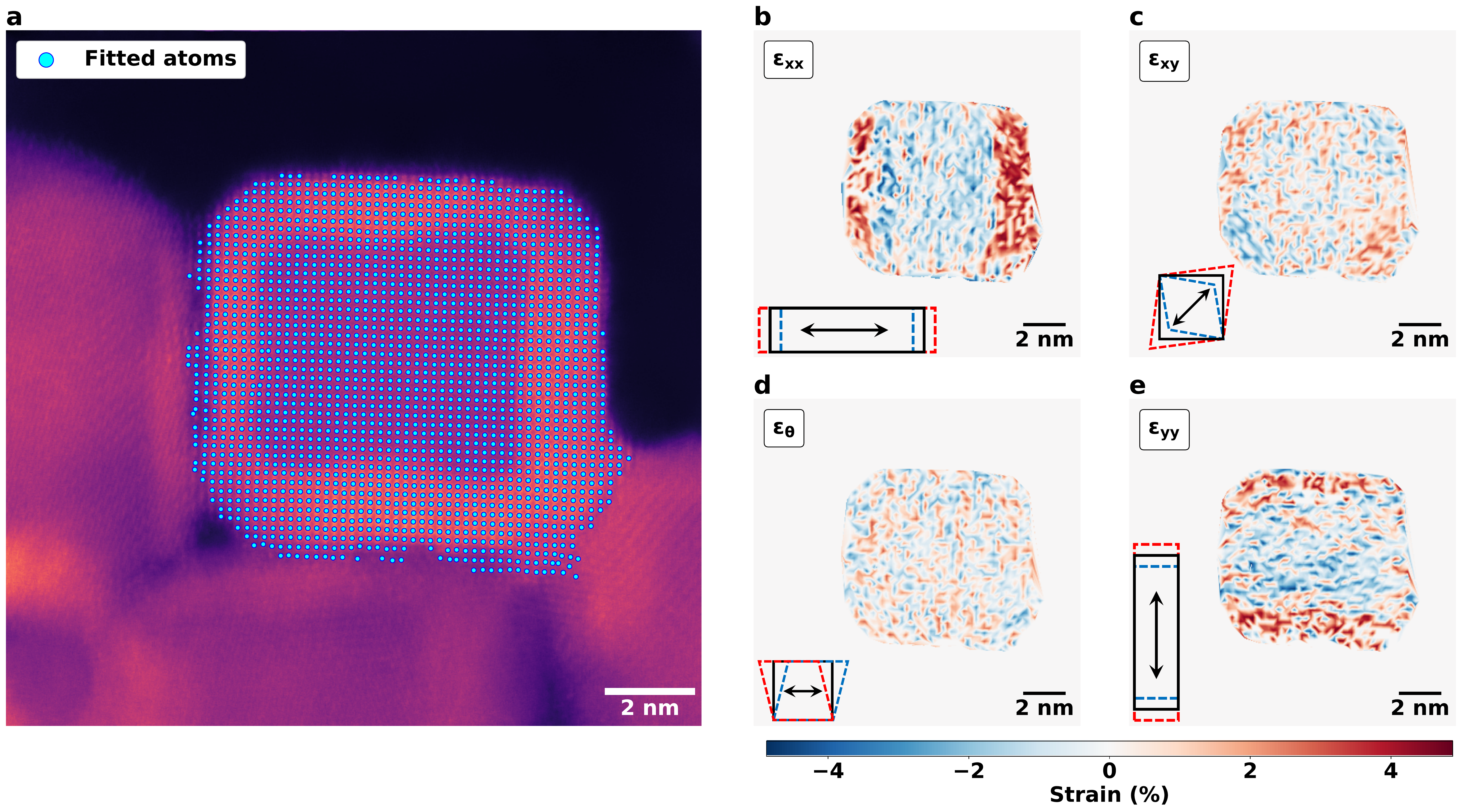}
	    	\caption{\label{fig:DC_Strain}\textbf{Lattice strain measurement from atomic resolution ADF--STEM images. (a)} Scan distortion corrected atomic resolution image of the Rh@Pt nanocube, with the refined atom positions overlaid as blue dots. \textbf{(b) - (e)} $\mathrm{\epsilon_{xx},\epsilon_{xy},\epsilon_{\theta}\ and\ \epsilon_{yy}}$ strain measured from the refined atom positions, with the graphically illustrated strain conventions shown in the bottom left of each individual strain map.}	
        \end{figure*}
        The advent of aberration correction has enabled sub-\r{a}ngstr\"{o}m resolution STEM imaging of atoms and atomic columns\cite{sub_angstrom}. When combined with Gaussian peak fitting, the assignment of the atom column positions with picometer precision can be performed\cite{pico_strain}. This approach has been used successfully to measure displacements at ferroelectric domain walls and to measure strain across interfaces\cite{nelson_DW,lno_DW}. In nanoparticle catalysts, for example, ADF-STEM imaging has been used to quantify strain by fitting the atom column positions with a precision below 1 pm or by performing geometric phase analysis (GPA) on the STEM images\cite{pico_strain,haadf_stem_NP_strain,nanoporous_gold_catalyst,pt_nanoparticle_stem,strain_coreshell_STEM}. 
	    
	    \autoref{fig:DC_Strain}(a) shows an ADF-STEM image of a Rh@Pt nanocube oriented along the $\mathrm{\left\langle 100\right\rangle}$ zone axis.  The ADF-STEM image is corrected for scan distortions\cite{drift_corr}, by collecting a pair of images with orthogonal scan directions (see \autoref{supp-fig:Strain00} and \autoref{supp-fig:Strain90} in the supplemental section for the individual images and strain measurements). Since the nanocube core consists of rhodium (atomic number = 45) and the shell consists of platinum (atomic number = 78), the shell exhibits more intense contrast than the core due to Z dependence of contrast in atomic resolution ADF-STEM imaging.\cite{z_contrast1,z_contrast2}. The atom columns are first identified as intensity maxima, and are then subsequently fitted as a 2-D Gaussian function\cite{mpfit}. The center of the Gaussian is the refined atom column position; the refined atom positions are overlaid on the ADF-STEM image in \autoref{fig:DC_Strain}(a) as blue dots. 
	    
	    Once the atom columns are assigned and located, the strain can be quantitatively determined by precisely measuring the column distance from its four orthogonal nearest neighbors. \autoref{fig:DC_Strain}(b) - (e) show the quantitative strain maps that result from applying this approach to the Rh@Pt nanocube in \autoref{fig:DC_Strain}(a) as $\mathrm{\epsilon_{xx}}$, $\mathrm{\epsilon_{xy}}$, $\mathrm{\epsilon_{\theta}}$ and $\mathrm{\epsilon_{yy}}$ respectively. The strain direction conventions used are overlaid on the individual images. We do not distinguish between the rhodium and platinum lattice in this strain quantification analysis and since the lattice constant of platinum $\mathrm{\left( a_{Pt} = 392.42pm \right)}$ is higher than that of rhodium $\mathrm{\left( a_{Rh} = 380.34pm \right)}$ by 3.17\%, both $\mathrm{\epsilon_{xx}}$ and $\mathrm{\epsilon_{yy}}$ in \autoref{fig:DC_Strain}(b) and \autoref{fig:DC_Strain}(e) respectively show approximately 4\% higher strain in the Pt shell as compared to the Rh core. Regions with 0\% strain in \autoref{fig:DC_Strain} thus have unit cells with a lattice spacing equivalent to that of unstrained Rh -- 380.34pm.
	    
	    Close inspection of the strain maps shows significant fluctuations in the measured strain that are visible as alternating high and low strained regions in the $\mathrm{\epsilon_{xx}}$ and $\mathrm{\epsilon_{yy}}$ maps. Notably, the direction of the strain fluctuation in $\mathrm{\epsilon_{xx}}$ strain is perpendicular to the fluctuation direction in $\mathrm{\epsilon_{yy}}$ strain. We also observe a cross-hatched pattern in $\mathrm{\epsilon_{xy}}$ and $\mathrm{\epsilon_{\theta}}$ strain maps. The question that arises is --- \textit{are these patterns real, or are they an artifact of scan distortions that could not be corrected?} Since the strain measurement depends on fitting multiple peaks and then measuring the relative inter-peak distances, the measurements will be sensitive to scan distortions unless extremely sophisticated image processing tools are applied to the image. Additionally, the absence of a reference distortion-free image makes it almost impossible to completely correct for distortions.
	    
	    Such problems associated with ADF-STEM quantification have been noted in other systems too. Yankovich \textit{et. al.}\cite{pico_strain}, Jones \textit{et. al.},\cite{Lewys_multiframe} and Savitzky \textit{et. al.}\cite{savitzky_drift} were able to bypass this issue by acquiring a large number of STEM images with short pixel dwell times, and then subsequently combining them to account for STEM scan distortions, or through the revolving STEM beam (RevSTEM) technique as demonstrated by Sang and LeBeau\cite{revstem}. GPA strain maps obtained from ADF-STEM by Daio \textit{et. al.}\cite{pt_nanoparticle_stem} demonstrate the challenges for understanding whether the variations in lattice parameter are a material-related phenomena or are a consequence of scan distortions.
	    
	    \subsection{\label{ssec:gpa}Measuring strain with Geometric Phase Analysis}
	    
	    \begin{figure*}
	    	\includegraphics[width=\textwidth]{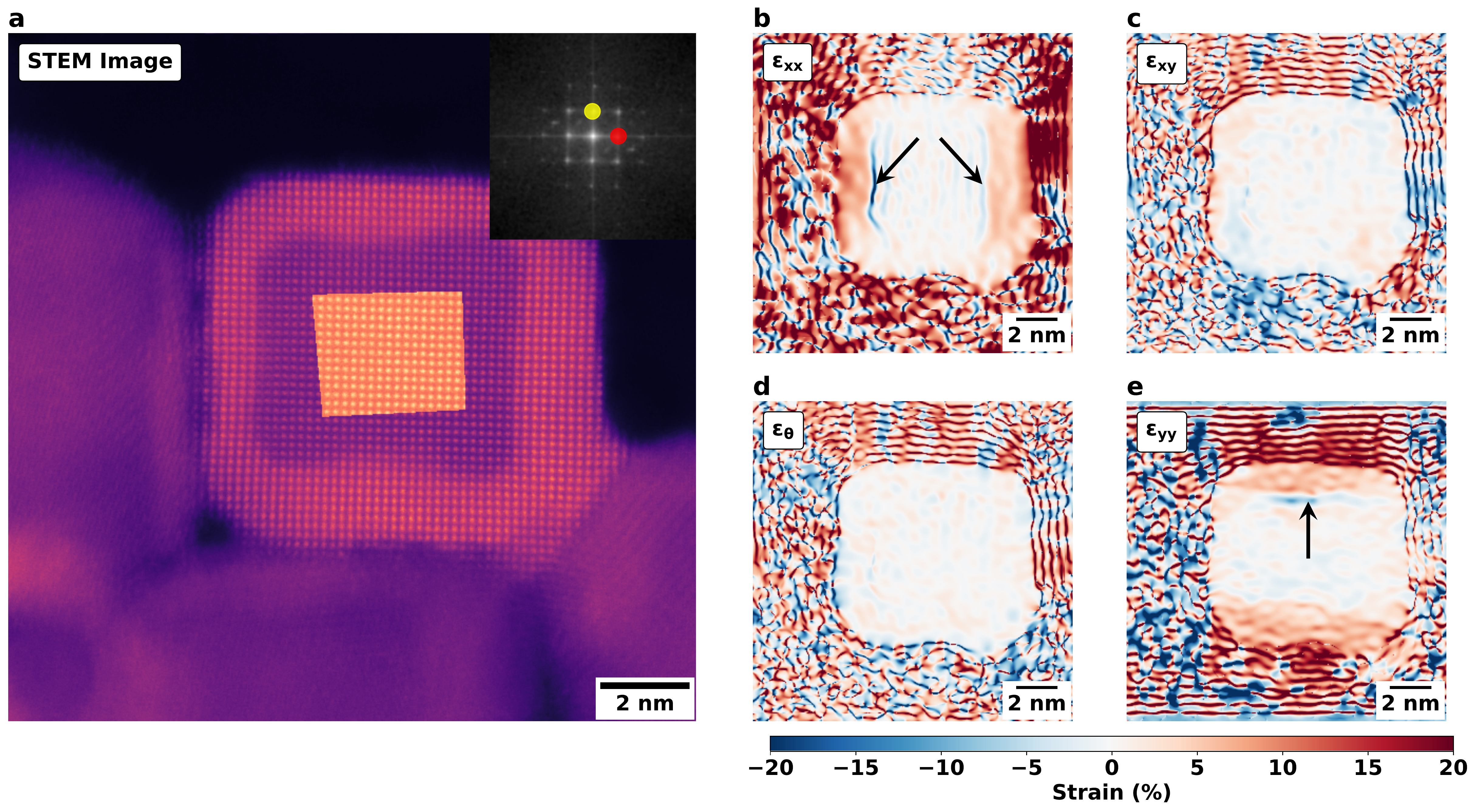}
	    	\caption{\label{fig:GPA}\textbf{Lattice strain measurement from geometric phase analysis (GPA) of atomic resolution ADF--STEM images. (a)} Scan distortion corrected atomic resolution image of the Rh@Pt nanocube, with the reference region marked by the parallelogram in the center of the cube. Inset: Fourier transform of the image, with the spots chosen for GPA analysis marked in yellow and red. \textbf{(b) - (e)} $\mathrm{\epsilon_{xx},\epsilon_{xy},\epsilon_{\theta}\ and\ \epsilon_{yy}}$ strain measured from GPA. The strain conventions followed are identical to the ones demonstrated in \autoref{fig:DC_Strain}(b) - \autoref{fig:DC_Strain}(e). The arrows point out compressive strain regions at the core-shell interface.}
	    \end{figure*}
    
    	Along with fitting atomic columns with two-dimensional Gaussians, we also pursued geometric phase analysis (GPA) on the atomic-resolution ADF-STEM datasets to quantify strain (\autoref{fig:GPA}). The principle behind geometric phase analysis is based on the idea that a translational variation in a real-space image is reflected as a phase variation in Fourier space. Thus, by comparing phase variations of non-colinear diffraction directions, lattice fluctuations and strain can be quantified from images. This is implemented through masked Fourier transforms of diffraction peaks, as demonstrated in the diffraction pattern in the inset of \autoref{fig:GPA}(a). If two such transforms could be obtained from non-colinear diffraction spots, then by comparing the variation of the \emph{phase} of the masked Fourier transforms, the lattice parameter variation can be tracked across an image\cite{original_GPA}.
       
       \autoref{fig:GPA}(b)-(e) map out the $\mathrm{\epsilon_{xx},\:\epsilon_{xy},\:\epsilon_{\theta}\:and\:\epsilon_{yy}}$ strain features respectively as calculated from GPA analysis on the scan distortion corrected ADF-STEM image, with striations in $\mathrm{\epsilon_{xx}}$ (\autoref{fig:GPA}(a)) and $\mathrm{\epsilon_{yy}}$ (\autoref{fig:GPA}(d)) demonstrating that scan distortion correction is unable to completely correct for scanning distortion effects. We additionally observe $\mathrm{\epsilon_{xx}}$ compressive strain at the core-shell interface in \autoref{fig:GPA}(b) as marked by the two arrows. Even though the underlying dataset remains unchanged in both atom position tracking and GPA analysis, similar $\mathrm{\epsilon_{xx}}$ compressive strain features are absent in \autoref{fig:DC_Strain}(b). This is attributed to artifacts in the GPA process itself in compound material interfaces\cite{GPA_problems}, rather than being an intrinsic material feature. Thus even though GPA is a widely used technique it is still not a error free strain measurement and can often paint misleading pictures of strain states -- showing strain fluctuations in areas where none may actually exist.
	 \subsection{\label{ssec:4D_STEM}4D-STEM imaging of core@shell nanocubes} 
	\begin{figure*}
		\includegraphics[width=\textwidth]{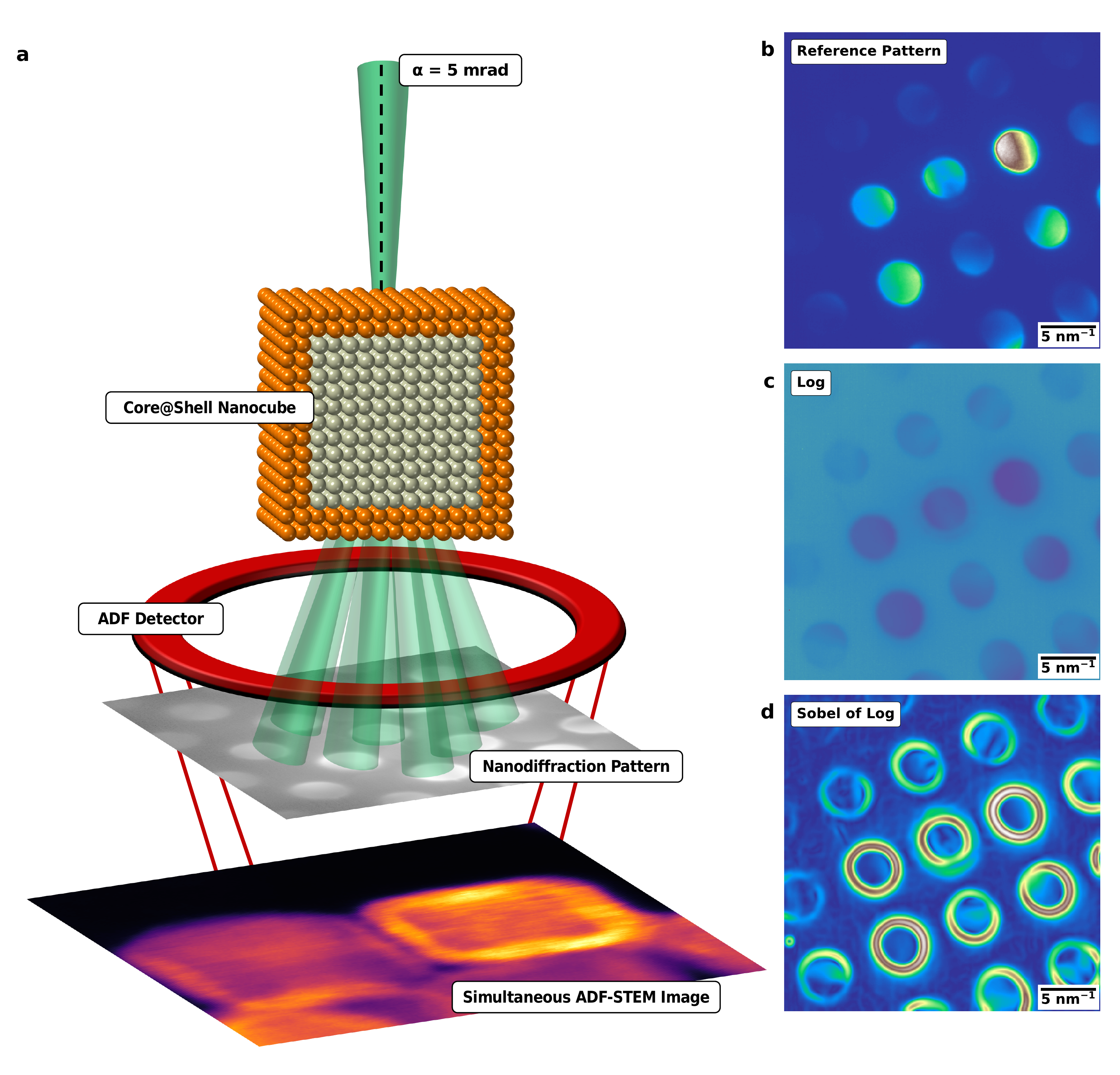}
		\caption{\label{fig:schematic}\textbf{4D-STEM experimental setup and data preconditioning. (a)} Schematic of the experimental setup with the electron probe semi-angle $\mathrm{\left( \alpha \right)}$ is 5mrad, ensuring that the diffraction disks do not overlap. An ADF detector present above the 4D-STEM camera also simultaneously captures an ADF image. \textbf{(b)} Raw reference CBED pattern without preconditioning. \textbf{(c)} Logarithm of the diffraction pattern in \autoref{fig:schematic}(b). \textbf{(d)} Magnitude of the Sobel filtered image of the logarithm of the diffraction pattern, shown in \autoref{fig:schematic}(c).}
	\end{figure*}
    To resolve problems that arise from quantifying strain from ADF-STEM images, we performed 4D-STEM imaging on the same Rh@Pt nanocube shown in \autoref{fig:DC_Strain}(a). In 4D-STEM experiments, rather than using an annular ring detector for ADF imaging or a circular detector used for bright field (BF) imaging, the full series of convergent beam electron diffraction (CBED) patterns arising from the beam-sample interactions is captured at every scan position. Four dimensions here refers to the 4D nature of the datasets obtained, where two dimensions correspond to the real space scanning coordinates and two dimensions correspond to the Fourier space diffraction patterns.
    
    The concept for measuring strain with 4D-STEM is to illuminate the sample with a unit-cell sized electron beam rather than a sub-\r{A} sized beam. This results in the so-called nanobeam electron diffraction (NBED), where the diffraction disks do not overlap with the central undiffracted transmitted electron beam, as shown schematically in \autoref{fig:schematic}(a) -- with the disk locations corresponding to the crystallographic axes as per Bragg's law. The higher order diffraction disk positions can then be compared to the disk location of the $\mathrm{\left\lbrace 000 \right\rbrace}$ transmitted beam, and the unit cell parameters perpendicular to the beam propagation direction can be ascertained at every scan position. Comparison of the unit cell with a reference unit cell allows the measurement of strain for that scanning pixel\cite{colin_4D_review,4d_stem_nbed_theory}. In atomic resolution STEM, strain is calculated first by fitting each individual atom columns which consist of a plural number of pixels, where each pixel is a scan position. Then these refined atom positions are compared to neighboring atoms. Since multiple scan locations are required to make a single strain estimate, this technique is susceptible to the effects of scan drift. Conversely, in 4D-STEM nanodiffraction, strain is measured from the individual diffraction pattern at a single scan position and thus it is less susceptible to scan distortion errors. This approach was first implemented on \textit{p}--doped MOSFET devices\cite{first_nbed_4D_strain}, and has been subsequently successfully applied to many different systems\cite{CMOS_4D_strain,direct_electron_4D_strain,insitu_4D_strain}. 
    
    4D-STEM has been used to quantify strain evolution with sub-picometer precision in monolayer \ce{WS2}--\ce{WSe2} heterostructures over a field of view (FOV) of hundreds of nanometers\cite{pico_4D_strain}. However there has been no standard and well-accepted routine for locating the diffraction disks to date. Han \textit{et. al.} used the center of mass (COM) of each diffraction disk to locate the disk positions with sub-pixel precision\cite{pico_4D_strain}. However theoretical simulations have demonstrated that the COM approach would fail to locate the disk positions accurately in thicker samples due to the presence of features within the diffraction disks\cite{4d_stem_nbed_theory}. Pekin \textit{et. al.} used several different approaches: Sobel filtering, cross-correlation and hybrid correlation to locate the diffraction disk positions, and observed significantly different strain distributions with different disk location approaches even when the underlying 4D-STEM dataset remained unchanged\cite{disk_registration}. A recent work from several of the same authors has attempted to circumvent this issue through using patterned condenser apertures, where a bulls-eye pattern is generated using focused ion beam (FIB), and have reported an order of magnitude improvement in disk location precision when using patterned apertures compared to unpatterned apertures\cite{patterned_aperture}. Rather than using hardware modifications, in this work we developed a data preconditioning routine by performing the Sobel filter operation on the logarithm of the CBED patterns (detailed in \autoref{supp-sec:precond} in the Supporting Information). Preconditioning the diffraction data has also recently been pursued through the power-cepstrum function to measure strain in nanoparticles from 4D-STEM NBED datasets, with the authors using cepstrum filtering to measure strain for multiple particles that may not always be oriented on a zone axis\cite{cepstrum}.
     
     \subsection{\label{ssec:strain_metrology}Strain metrology from preconditioned 4D-STEM datasets}
     	\begin{figure*}
     	\includegraphics[width=\textwidth]{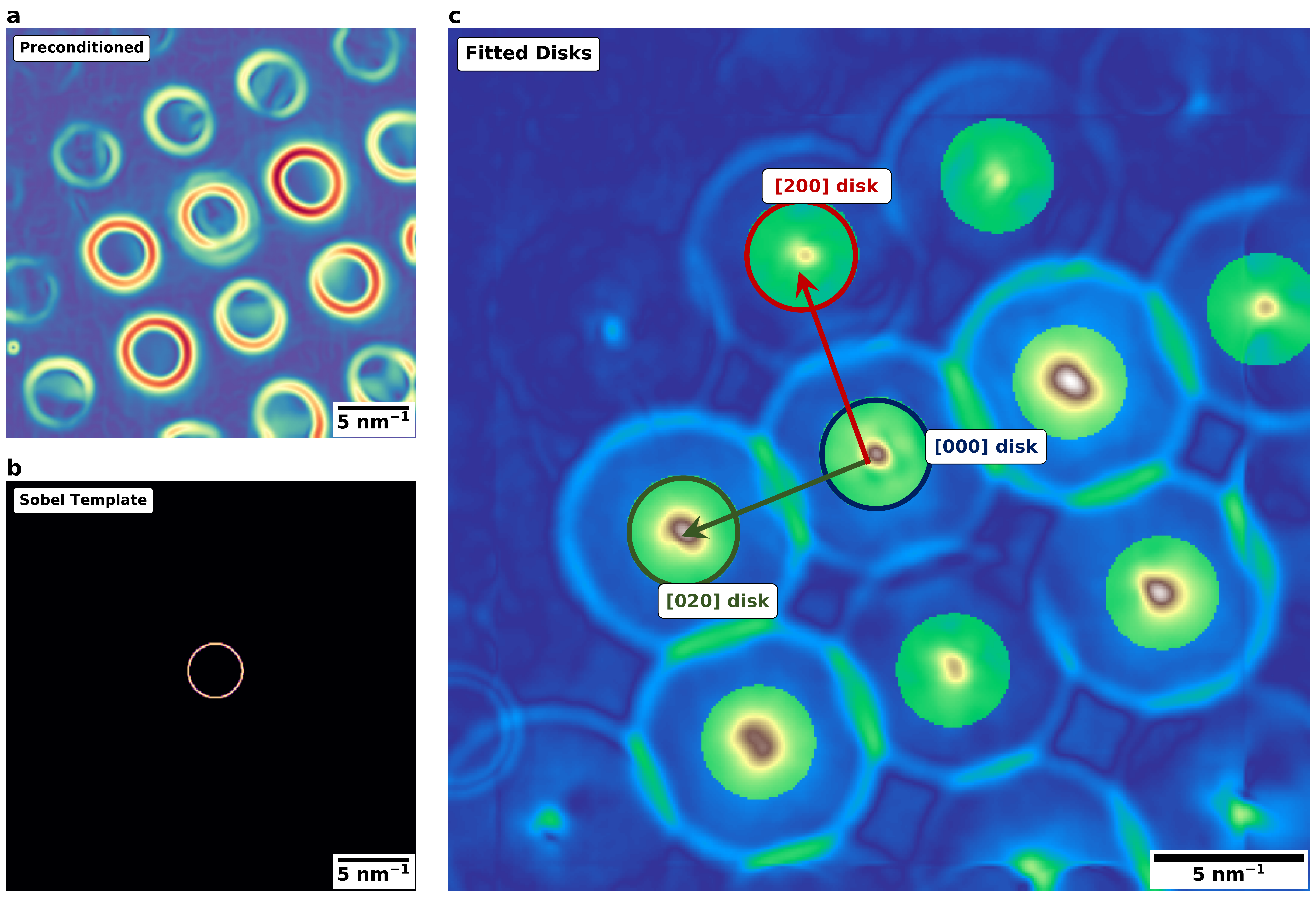}
     	\caption{\label{fig:disk_fit}\textbf{Locating disk positions from pre-conditioned data. (a)} Preconditioned CBED pattern. \textbf{(b)} Sobel magnitude of the template used for cross-correlation. \textbf{(c)} Cross-correlation of the \autoref{fig:disk_fit}(a) with \autoref{fig:disk_fit}(b), with the located disk positions overlaid. Two of the disks --  $\mathrm{\left[ 020 \right]}$ and $\mathrm{\left[ 200\right]}$ are highlighted showing the vectoral distance from the undiffracted $\mathrm{\left[ 000 \right]}$ disk}
     \end{figure*} 
    
    While patterned apertures are an extremely promising avenue for strain quantification, they require sophisticated hardware modifications that sometimes may not be possible in a general purpose imaging equipment. The core idea of aperture patterning is to overlay a common feature in each of the diffraction disks. This increases the similarity between the disks and consequently improves the precision of the cross-correlation. We use this same philosophy in our own data preconditioning routine. Since each diffraction disk must have the same size in Fourier space as the central transmitted disk, thus the most common parameter for every disk is the disk edge rather than the intensity or the features inside the disk. We exploit this idea in our preconditioning routine by first obtaining a logarithm of the CBED pattern (\autoref{fig:schematic}(c)) to damp out the features inside the disk followed by Sobel filtering of the logarithm data to generate the disk edge (\autoref{fig:schematic}(d)).
 
    The preconditioned CBED patterns at every scanning pixel (a single pattern is shown in \autoref{fig:disk_fit}(a)) are then subsequently cross-correlated with the edge of a diffraction disk (the Sobel template shown in \autoref{fig:disk_fit}(b)), with the result from the cross-correlation for a single CBED pattern shown in \autoref{fig:disk_fit}(c). Each disk location is now replaced with a sharp peak, which is fitted with a 2D Gaussian function to locate the peak position with sub-pixel precision in the diffraction space (see \autoref{supp-fig:Filter} in the Supporting Information for a comparison of the peak sharpness between conditioned and raw data). 
    
    For the Rh@Pt nanocube investigated in this work, nine peak positions were located for every single CBED pattern, as shown in \autoref{fig:disk_fit}(c), with the central peak referring to the $\mathrm{\left\lbrace 000 \right\rbrace}$ transmitted electron beam, as marked by the blue circle in \autoref{fig:disk_fit}(c). The other peaks correspond to the higher order diffraction planes, with the peaks corresponding to the $\left( 020 \right)$ and the $\mathrm{\left( 200 \right)}$ diffraction planes marked with green and red circles respectively in \autoref{fig:disk_fit}(c). Once the peak positions are determined with sub-pixel precision by fitting a 2D Gaussian to the observed peak, the vectoral distance of each higher diffraction disk from the central transmitted $\mathrm{\left\lbrace 000 \right\rbrace}$ disk is measured, thus giving eight inverse inter-planar spacings for the pattern under investigation. The distances corresponding to the inverse of $\mathrm{\left( 020 \right)}$ and the $\mathrm{\left( 200 \right)}$ inter-planar spacings are visualized by the green and red arrows respectively in \autoref{fig:disk_fit}(c). The distances measured in the CBED pattern are inverse of the real-space parameters since the diffraction pattern corresponds to the Fourier transform of the convolution between the electron beam and the crystal being imaged. The measured inter-planar spacings from a CBED pattern are subsequently used to solve Bragg's equation for the unit cell. Thus, the unit cell parameters corresponding to the region of the crystal illuminated by the electron beam \emph{at a single scan position} is calculated from each individual CBED pattern. Since a single CBED pattern is sufficient for calculating the unit cell at that scan position, the measured strain in 4D-STEM is independent of scan distortions, unlike in ADF-STEM imaging.
    
    \begin{figure*}
    	\includegraphics[width=\textwidth]{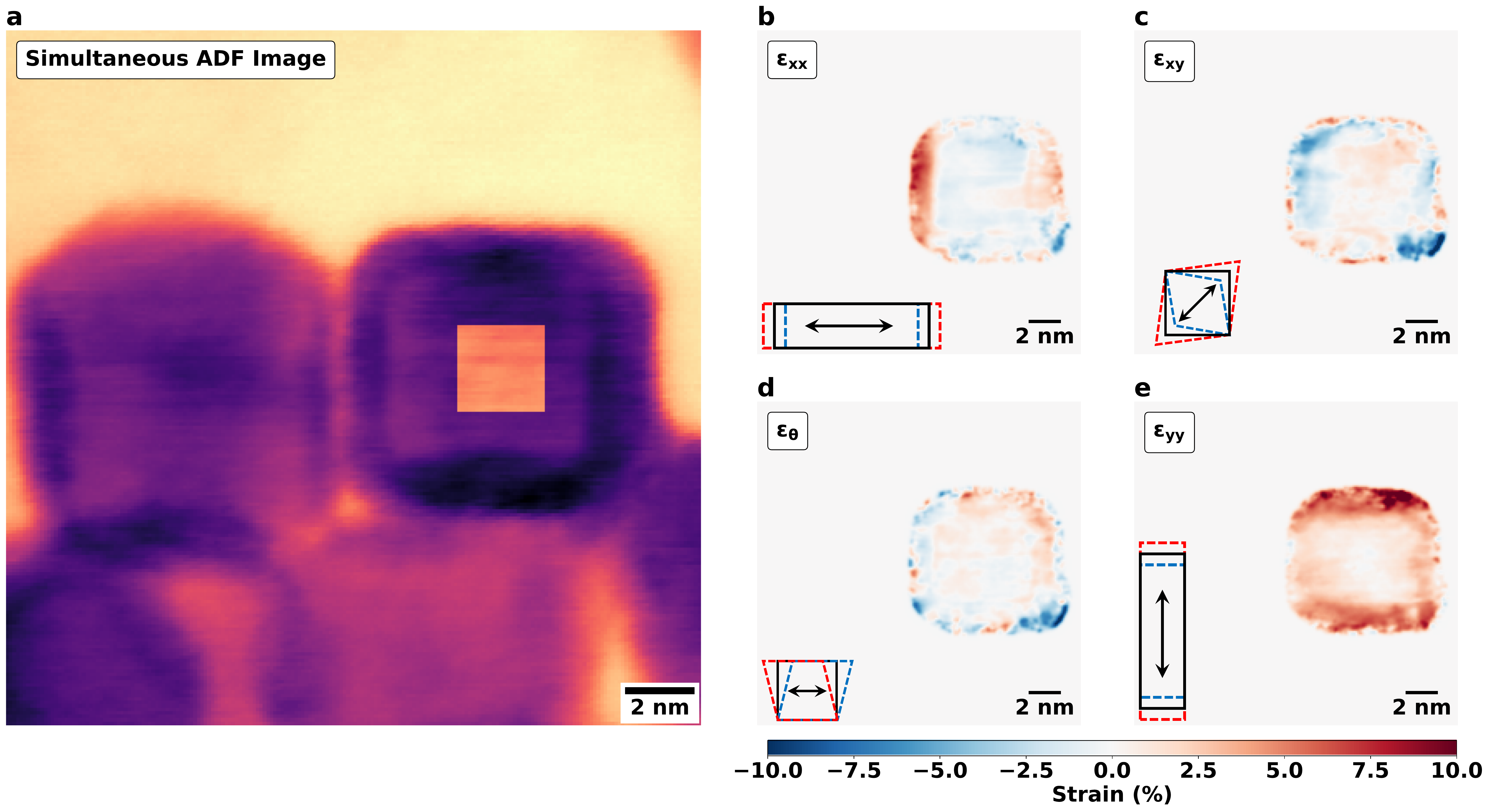}
    	\caption{\label{fig:strain_map}\textbf{Strain measured from preconditioned 4D-STEM data. (a)} Simultaneously acquired non atomic resolution ADF-STEM image with the reference region overlaid. \textbf{(b) - (e)} $\mathrm{\epsilon_{xx},\epsilon_{xy},\epsilon_{\theta}\ and\ \epsilon_{yy}}$ strain respectively in the nanocube compared to the averaged unit cell from the reference region in \autoref{fig:strain_map}(a), with the graphically illustrated strain conventions shown in the bottom left of each individual strain map.}
    \end{figure*}
    
    The unit cell parameters calculated from the CBED pattern only have four unique terms; however, since the parameters along the beam propagation direction cannot be measured from zero order Laue zone (ZOLZ) peak locations and thus the strain measured is also two-dimensional and measures only an averaged strain perpendicular to the electron beam propagation vector. 
    
    Unlike microscopes with physical apertures, NION microscopes have a lens-defined virtual aperture, and thus the exact aperture sizes are not known. Thus in this work, we measure strain by comparing the unit cell calculated at every scan position with a reference unit cell. \autoref{fig:strain_map}(a) demonstrates the reference region selected from the center of the nanocube core. The mean preconditioned CBED pattern is calculated for this reference region and then the unit cell is calculated for this mean pattern. At every scan point, the calculated unit cell is then compared to the unit cell from the reference region, and the strain is subsequently calculated based on the formula originally given by Pekin \textit{et. al.}\cite{disk_registration}. Similar to strain maps obtained from ADF-STEM images through atom position analysis (see \autoref{fig:DC_Strain}) or GPA analysis (see \autoref{fig:GPA}), we find approximately 5\% higher $\mathrm{\epsilon_{xx}}$ along the x direction (\autoref{fig:strain_map}(b)), since the platinum shell has a larger unit cell size. This is also observed in the $\mathrm{\epsilon_{yy}}$ strain map (\autoref{fig:strain_map}(e)), where the strain is approximately 5\% higher along the y direction. Unlike strain maps obtained from atomic resolution datasets, striated variations in the strain are not visible - indicating that the features observed in \autoref{fig:DC_Strain}(b) - \autoref{fig:DC_Strain}(e) were an artifact of scanning distortions rather than being a property of the nanocube itself., as compared to \autoref{supp-fig:Strain00} and \autoref{supp-fig:Strain90}. 
    
    \subsection{\label{ssec:MCR}Identifying regions with Multivariate Curve Resolution} 
    \begin{figure*}
    	\includegraphics[width=\textwidth]{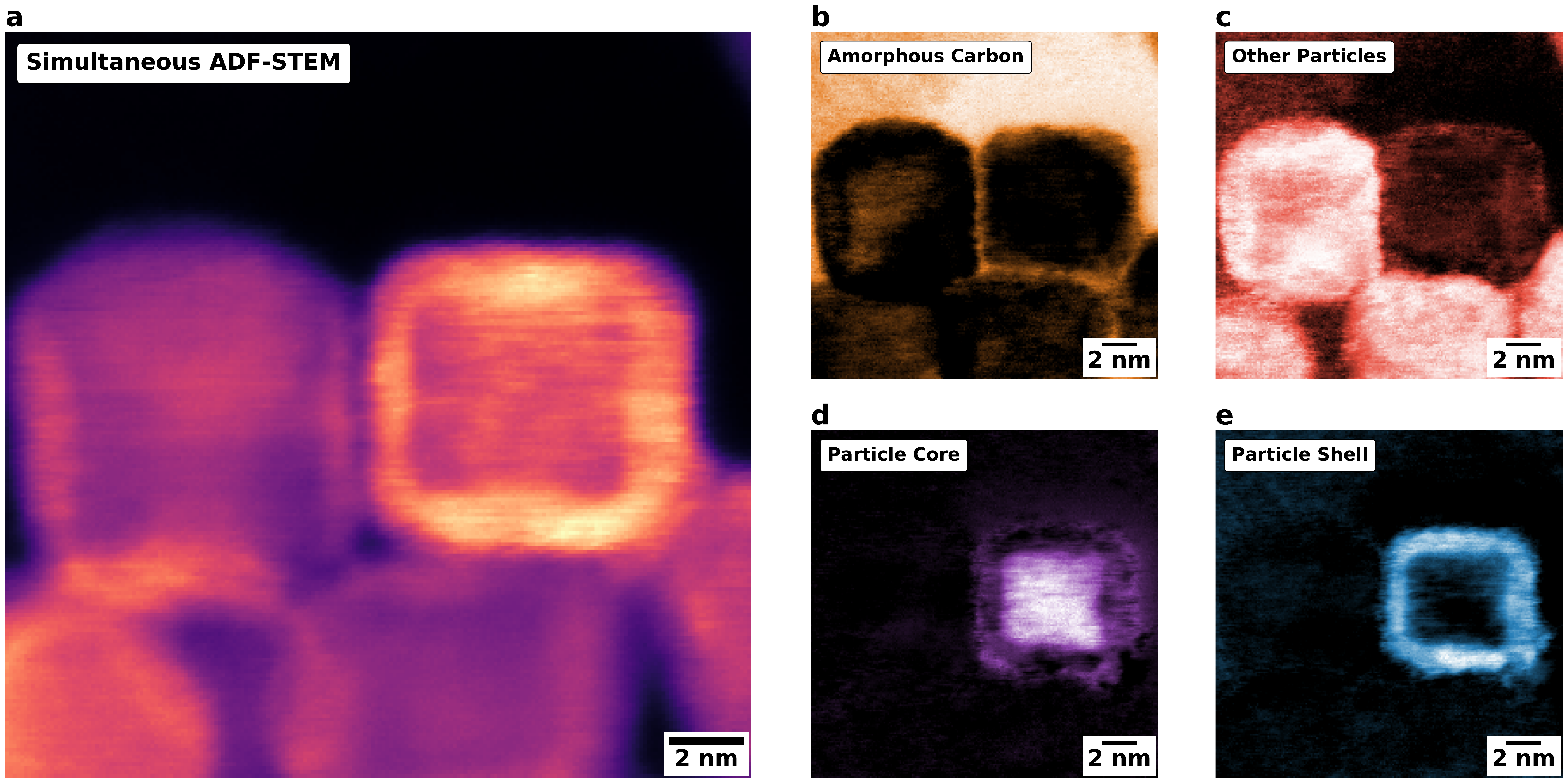}
    	\caption{\label{fig:MCR}\textbf{Identification of regions with multivariate curve resolution (MCR) on preconditioned data. (a)} Simultaneously collected ADF-STEM image, with the microscope electron beam convergence angle at 5mrad. \textbf{(b)- (e)} Amorphous carbon, neighboring nanocubes, Rh core and Pt shell regions shown respectively, as calculated from multivariate curve resolution.}
    \end{figure*}
    One of the most distinguishing features of ADF-STEM imaging is the effect of \textit{Z-contrast}, where the intensity of the atomic columns is proportional to the total atomic number (Z) of the atoms comprising the column being imaged\cite{z_contrast1,z_contrast2}. This makes identification of core@shell structures exceptionally simple, as demonstrated in \autoref{fig:DC_Strain}(a) where the platinum shell is brighter than the rhodium core. While this Z dependence of contrast is still visible in \autoref{fig:MCR}(a), where the shell is brighter than the core, the accurate identification and assignment of regions being imaged is still challenging in 4D-STEM data sets. For core@shell nanoparticles there is an added complexity, since the FOV often also includes other nanoparticles which may be misoriented with respect to the microscope's optic axis and also the presence of the amorphous carbon support film. To determine the region of interest (ROI) for strain mapping from the FOV, we performed multivariate curve resolution (MCR) on the 4D-STEM datasets. 
    
    MCR is a technique for calculating the concentrations of individual pure spectral signatures at each acquisition point from a mixed signal. For example, if a spectroscopic signal is obtained with respect to time, and at every point of time there are contributions from multiple pure individual spectra, MCR will generate the relative contribution of each pure spectra with time, and is therefore often referred to as \textit{spectral unmixing} or \textit{endmember extraction}\cite{MCR,MCR_ALS}. Multiple different iteration schemes can be used for unmixing in MCR, with alternating least squares (ALS) being the most commonly used. For this work, we used the \texttt{pyMCR} routine which uses the alternating regression (AR) scheme\cite{pymcr,pymcr_journal}.
    
    To identify the regions in our scanning image, we chose the flattened NBED pattern at each scan position as the spectra to be unmixed, and we use the the flattened diffraction patterns from amorphous carbon, nanocubes etc. as the individual pure spectral signatures. Multiple different techniques have also been proposed for estimating the number of unique pure spectral signatures and also the individual pure spectra themselves -- with singular value decomposition and principal component analysis being the most commonly used computational methods. For our work, reference pure spectral signatures were chosen by manually locating the different regions of the sample (the amorphous carbon, neighboring nanocubes, Rh core and Pt shell) from the simultaneous ADF-STEM image and then taking the average of the flattened diffraction pattern from each manually assigned region (see \autoref{supp-fig:MCR_ROI} in the Supporting Information for manual ROIs used for calculating the pure spectra). We performed MCR analysis of the original unprocessed data, the log of the data and the log-Sobel filtered data and observed the best results with the log-Sobel filtered 4D datasets (see \autoref{supp-fig:MCR_Regions_Raw},\autoref{supp-fig:MCR_Regions_Log} and \autoref{supp-fig:MCR_Regions} in the Supporting Information for a comparison of the concentration profiles). 
    
    \autoref{fig:MCR}(b)-\autoref{fig:MCR}(e) show visualizations of the different regions of the 4D data assigned by the MCR algorithm. We found MCR to be suitable not only for distinguishing the nanocube from the neighboring nanocubes (\autoref{fig:MCR}(c)) and amorphous carbon (\autoref{fig:MCR}(b)), but also for distinguishing between the Rh core (\autoref{fig:MCR}(d)) and the Pt shell (\autoref{fig:MCR}(e)). Using the data from MCR we can therefore assign scan regions as belonging to either the nanocube core or the shell, and perform a comparison of the strain between the two regions, and measure the evolution of strain and unit cell size across the core@shell interface.
    
    \subsection{\label{ssec:UC_Size}Unit cell size variation in the nanocube} 
	\begin{figure*}
		\includegraphics[width=\textwidth]{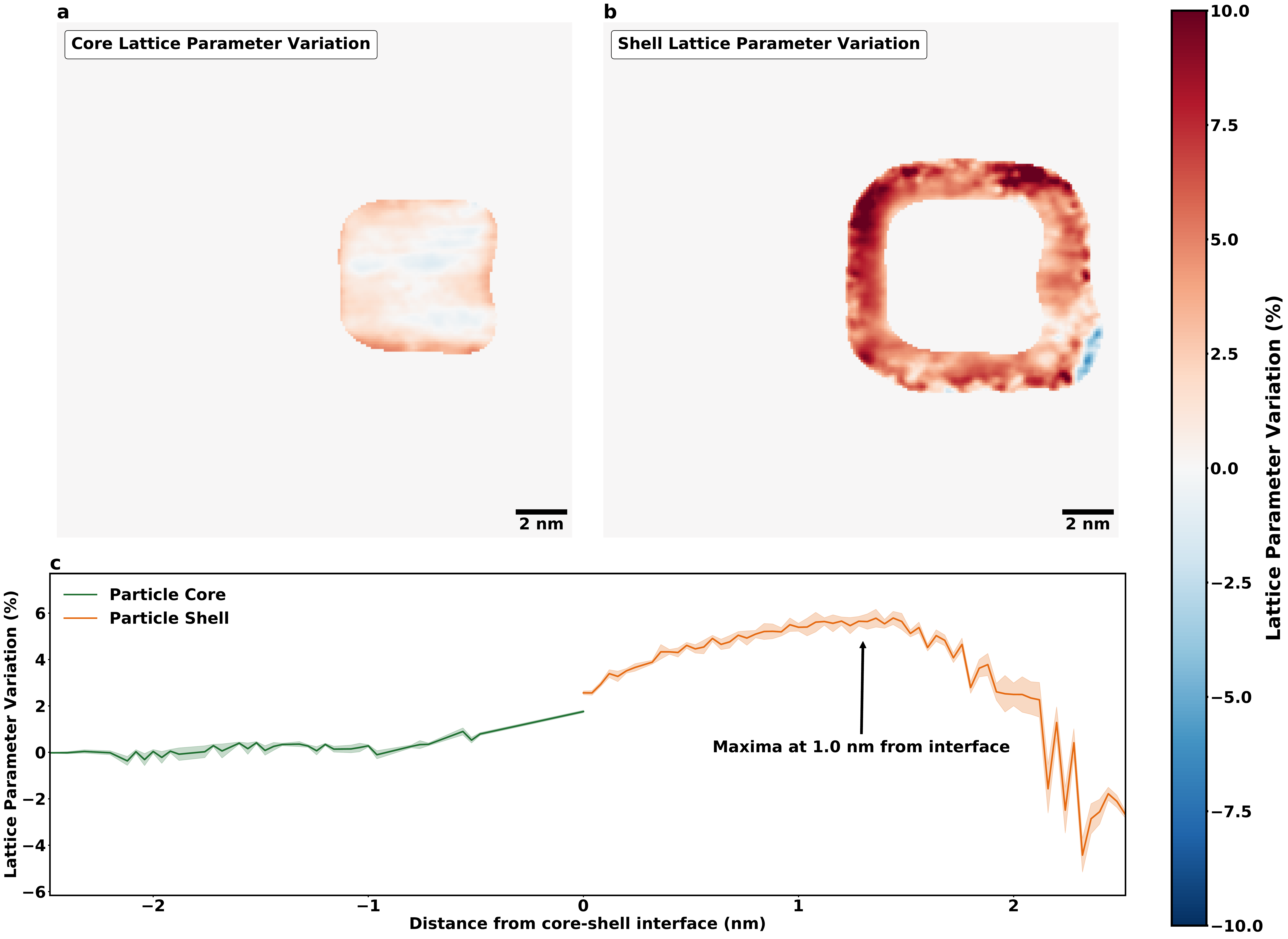}
		\caption{\label{fig:UC_map_4D}\textbf{Evolution of unit cell size in the nanocube core and shell. (a)} Lattice parameter variation in the \textit{Rh core} compared to the reference region in \autoref{fig:strain_map}(a). \textbf{(b)} Lattice parameter variation in the \textit{Pt shell} compared to the reference region in \autoref{fig:strain_map}(a). \textbf{(c)} Evolution of the change in lattice parameter as a function of the distance from the core@shell interface. The lattice parameter is measured with respect to the lattice parameter of the reference region in \autoref{fig:strain_map}(a).}
	\end{figure*}
    
   The accuracy of mapping strain from cross-validation on preconditioned datasets is approximately 0.07\% (see Supporting Information) --- which enables sub-picometer precision strain and unit cell size measurements. When combined with MCR, we can assign calculated unit cells to either the nanocube core or the shell, thereby enabling a direct comparison. \autoref{fig:UC_map_4D}(a) maps the unit cell variation in the Rh core when compared to the reference unit cell (\autoref{fig:strain_map}(a)), while \autoref{fig:UC_map_4D}(b) maps it in the Pt shell. In \autoref{fig:UC_map_4D}(a) we observe that the unit cell size is not uniform in the core and increases as we move towards the Rh@Pt interface. This indicates that epitaxial growth of a shell whose lattice is mismatched with the core also strains the core. In \autoref{fig:UC_map_4D}(b) we observe that the lattice parameter of the shell is not uniform all throughout, indicating a more complex picture of strain than that indicated by simplistic lattice mismatch models. 
   
   \autoref{fig:UC_map_4D}(c) plots the mean lattice parameter of the nanocube as a function of the distance from the core-shell interface, with the error  of the measurements calculated as $\mathit{^{\sigma}/_{\sqrt{n}}}$, where $\mathit{\sigma}$ is the standard deviation of the measured lattice parameter, and $\mathit{n}$ is the number of measurements. We observe that the unit cell size (plotted in green) increases in the core as we approach the core@shell interface, as \autoref{fig:UC_map_4D}(a) also shows. In the Rh shell, plotted in orange in  \autoref{fig:UC_map_4D}(c), however the lattice parameter actually reaches a maxima - located 1 nm from the core@shell interface, and then decreases as we approach the Pt surface. The value reached at this maxima is $\mathrm{\approx 6\%}$ higher when compared to the reference unit cell, while the difference in the lattice parameter between relaxed rhodium and platinum is 3.17\%. The difference is higher in our experiments since it is likely that the reference region in the core itself is compressively strained. Interestingly, molecular dynamics simulations from atomistic models of these core@shell Rh@Pt nanocubes reveal compressively strained cores\cite{CDI,josie_sara}. Additionally, the presence of a maxima in the unit cell size in the nanocube core suggests that the surface rearrangement of atoms lead to compressive stresses. Thus plotting the unit cell variations we can see a significantly more complex picture of strain, that is only visible because of the higher precision and absence of scan distortions that are afforded by 4D-STEM in contrast to aberration-corrected atomic resolution STEM imaging. Considering the case of core@shell Rh@Pt nanocubes further, an understanding of the complex strain distribution is essential to identifying accurately how structure contributes to performance. A series of similarly sized core@shell Rh@Pt nanocubes with different Pt shell thicknesses were previously prepared and evaluated as electrocatalysts for formic acid oxidation, where a volcano-like dependence on performance was reported as a function of shell thickness\cite{skrabalak_RhPt}. This trend was expected, but the reported decrease in CO poisoning at a Pt shell thickness of $\approx$6 monolayers (and corresponding enhanced performance) was unexpected as the surface is anticipated to be relaxed at such thicknesses. The approach taken here provides the necessary precision to identify variations within individual nanoparticles (e.g., at corners versus faces) that may account for such discrepancies but a full analysis of the original dataset is beyond the scope of this manuscript.
    
    \subsection{\label{ssec:Compare}Comparison of strain metrology techniques}
    
    \begin{figure*}
    	\includegraphics[width=\textwidth]{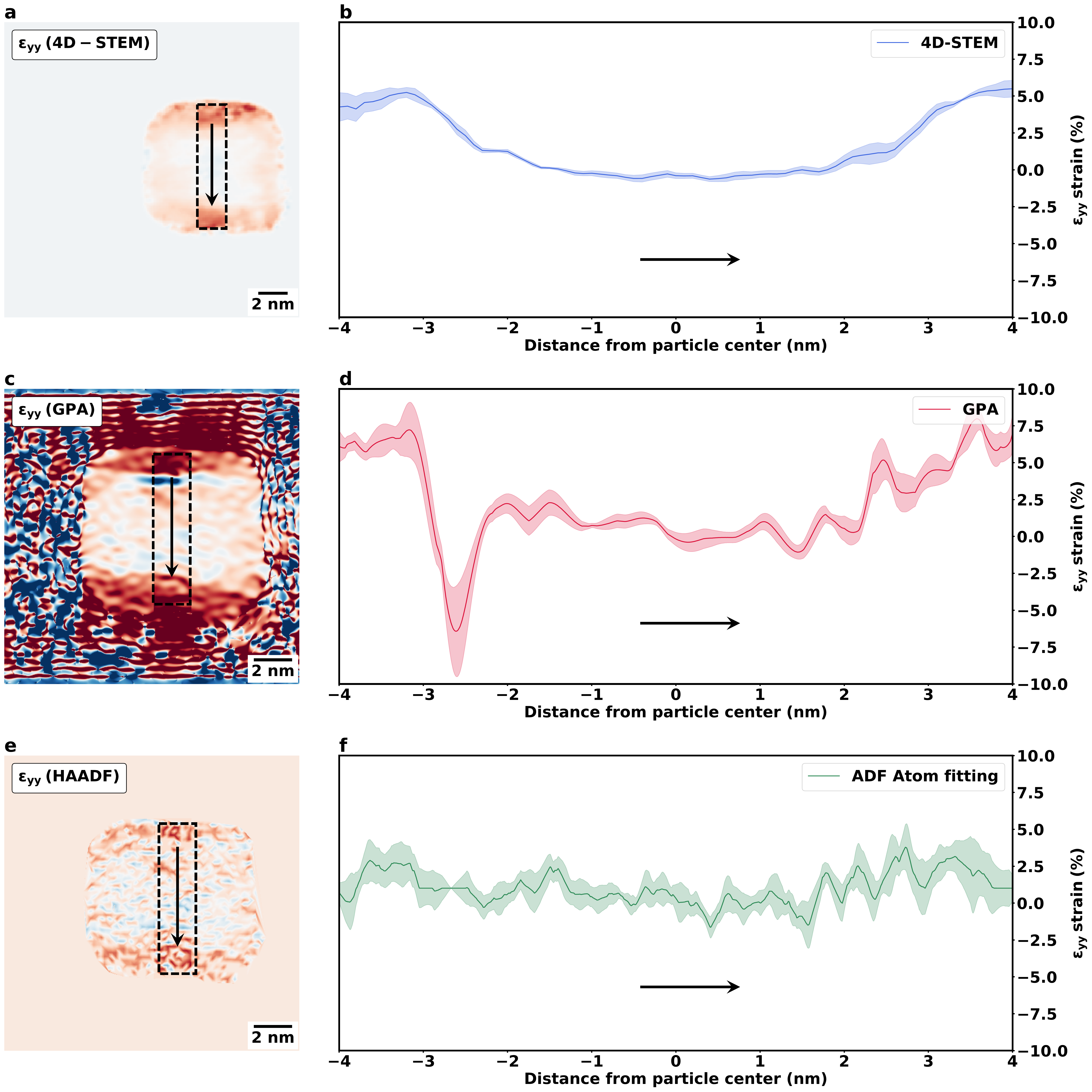}
    	\caption{\label{fig:strain_compare}\textbf{Comparision of measured strain through multiple techniques. (a)} $\mathrm{\epsilon_{yy}}$ strain measured through 4D-STEM disk fitting. \textbf{(b)} Variation of $\mathrm{\epsilon_{yy}}$ strain in the region marked by the black dashed rectangle in \autoref{fig:strain_compare}(a) along the arrow direction. \textbf{(c)} $\mathrm{\epsilon_{yy}}$ strain measured through geometric phase analysis. \textbf{(d)} Variation of $\mathrm{\epsilon_{yy}}$ strain in the region marked by the black dashed rectangle in \autoref{fig:strain_compare}(c) along the arrow direction.. \textbf{(e)} $\mathrm{\epsilon_{yy}}$ strain measured through two dimensional gaussian fitting of individual atom columns in a HAADF-STEM image. \textbf{(f)} Variation of $\mathrm{\epsilon_{yy}}$ strain in the region marked by the black dashed rectangle in \autoref{fig:strain_compare}(e) along the arrow direction.}
    \end{figure*}

    In this work, we have used three different techniques - atom column tracking, geometric phase analysis and 4D-STEM nanodiffraction to quantitatively measure strain in the \textit{same} nanoparticle. While visual examination shows that the strain maps are similar, we tested the same section of the measured $\mathrm{\epsilon_{yy}}$ strain map to compare between the techniques.  \autoref{fig:strain_compare}(a) shows the 4D-STEM strain map while  \autoref{fig:strain_compare}(b) shows the strain in the section marked by the black dashed rectangle in  \autoref{fig:strain_compare}(a). The strain is plotted along the direction of the arrow in both the pictures, and is averaged in the direction perpendicular to the arrow, with the shaded region in the plot being the variance of the strain. Similarly  \autoref{fig:strain_compare}(c) and  \autoref{fig:strain_compare}(d) demonstrate the average $\mathrm{\epsilon_{yy}}$ strain as measured through GPA, while \autoref{fig:strain_compare}(e) and  \autoref{fig:strain_compare}(f) demonstrate strain for atom column tracking. Since the same region of the same particle is chosen for comparison, we would expect close to identical plots in \autoref{fig:strain_compare}(b), \autoref{fig:strain_compare}(d) and \autoref{fig:strain_compare}(f). While \autoref{fig:strain_compare}(b) demonstrates that the shell has a higher lattice parameter as compared to the core, GPA plots in \autoref{fig:strain_compare}(d) show a localized region of compressive strain, which is not present in the 4D-STEM data. As explained previously, this compressive strain is not a real feature, but is rather an artifact of the GPA method itself\cite{GPA_problems}. Conversely, in the averaged strain measured from atom column tracking (\autoref{fig:strain_compare}(f)) it is almost impossible to visualize the higher unit cell parameter in the particle shell as compared to the core, even when it can be ascertained visually from the $\mathrm{\epsilon_{yy}}$ strain map in \autoref{fig:strain_compare}(e). The variance of the strain is also significantly higher in both the GPA and atom column tracking data as compared to the 4D-STEM strain data.
    
    While the importance of strain in modifying catalytic activity is well known, \autoref{fig:strain_compare} shows how error prone STEM based metrology techniques can be. GPA can show compressive strain in regions where non exist, while scan distortion effects can wash out features in atom tracking methods.
    	    
    \section{\label{sec:conclusions}Conclusions}
    
    In this work, we demonstrate the utility of 4D-STEM to quantitatively measure strain in core@shell catalyst nanocubes. Our results indicate that the picture of strain is significantly more complicated in Rh@Pt core@shell nanocubes than simply due to lattice mismatch dictating the unit cell size in the shell. We also demonstrate that preconditioning the 4D-STEM nanobeam electron diffraction datasets allows the precise identification of the core and shell atom positions using MCR. Performing disk location analyses on the preconditioned data additionally enables sub-picometer precision strain measurements without the detrimental effects of drift distortions. Two features that are within the noise in ADF-STEM measurements are clearly visible in 4D-STEM measurements; the nanocube core does not have a consistent unit cell size with the cell size increasing as it approaches the core -- shell interface and two, the unit cell size in the Pt shell reaches a maxima that is between the nanocube surface and Rh@Pt interface.
    
    Our results and techniques developed here thus allow for high precision strain measurements across interfaces and allow quantitative estimations of the effect of interfaces on strain. This is a technique that can be extended beyond nanoparticles too into other systems such as semiconductor heterojuctions, thin films, ferroelectric domains, etc. Additionally, the strain results point to a much more complex picture for core@shell nanoparticles. The unit cell size of the shell is not constant and \textit{d-band} engineering through epitaxy needs to take into account surface effects and shell thicknesses. The core is not unaffected by the shell and undergoes both compressive and tensile strain depending on its distance from the core@shell interface. Future work on strain engineered nanoparticles must take into account these complexities for developing highly active electrocatalysts.
    
    \section{Author Contributions}
    
    D.M. and R.R.U. designed the study. J.T.L.G. and S.E.S. prepared the core@shell nanocubes and transferred them to TEM grids. D.M. performed the high-resolution STEM and 4D-STEM experiments. D.M. developed the Python routines for analyzing the datasets, analyzed the experimental data and wrote the paper. All authors discussed the results and commented on the manuscript.
    
    \section{Conflicts of Interest}
    
    The authors declare no conflicts of interest.
    
	 This manuscript has been authored by UT-Battelle, LLC under Contract No. DE-AC05-00OR22725 with the U.S. Department of Energy. The United States Government retains and the publisher, by accepting the article for publication, acknowledges that the United States Government retains a non-exclusive, paid-up, irrevocable, world-wide license to publish or reproduce the published form of this manuscript, or allow others to do so, for United States Government purposes. The Department of Energy will provide public access to these results of federally sponsored research in accordance with the DOE Public Access Plan (\href{http://energy.gov/downloads/doe-public-access-plan}{http://energy.gov/downloads/doe-public-access-plan}).
    
     \section{Code and Data Availability}
     
     The Python codes for analysis are available on GitHub\cite{stemtools}. Experimental ADF-STEM and 4D-STEM datasets and Jupyter notebooks used for the analysis will be made available upon publication. 
     
     \begin{suppinfo}
     	Effects of scanning distortions on strain measurement,data preconditioning routines, manual ROIs for MCR analysis, effect of preconditioning on MCR analysis.  	
     \end{suppinfo}
    
    \section{Acknowledgments}
    
    This research was supported by ORNL’s Laboratory Directed Research and Development (LDRD) Program, which is managed by UT-Battelle, LLC for the U.S. Department of Energy (DOE) (D.M. and R.R.U.). J.T.L.G. and S.E.S. were supported by U.S. DOE BES Award DE-SC0018961. Electron microscopy was conducted as part of a user proposal at Oak Ridge National Laboratory’s Center for Nanophase Materials Sciences (CNMS), which is a U.S. DOE Office of Science User Facility. D.M. and R.R.U. would like to acknowledge the use of resources of the Compute and Data Environment for Science (CADES) at the Oak Ridge National Laboratory, which is supported by the Office of Science of the U.S. DOE under Contract No. DE-AC05-00OR22725. We would like to thank Michael Zachman and Jordan Hachtel for helpful discussions. D.M. would like to thank Andrew Lupini for help with setting up the nanodiffraction imaging on the NION microscopes.
    
	\bibliography{manuscriptReview}
	
	\begin{tocentry} 
		\includegraphics[width=3.35in]{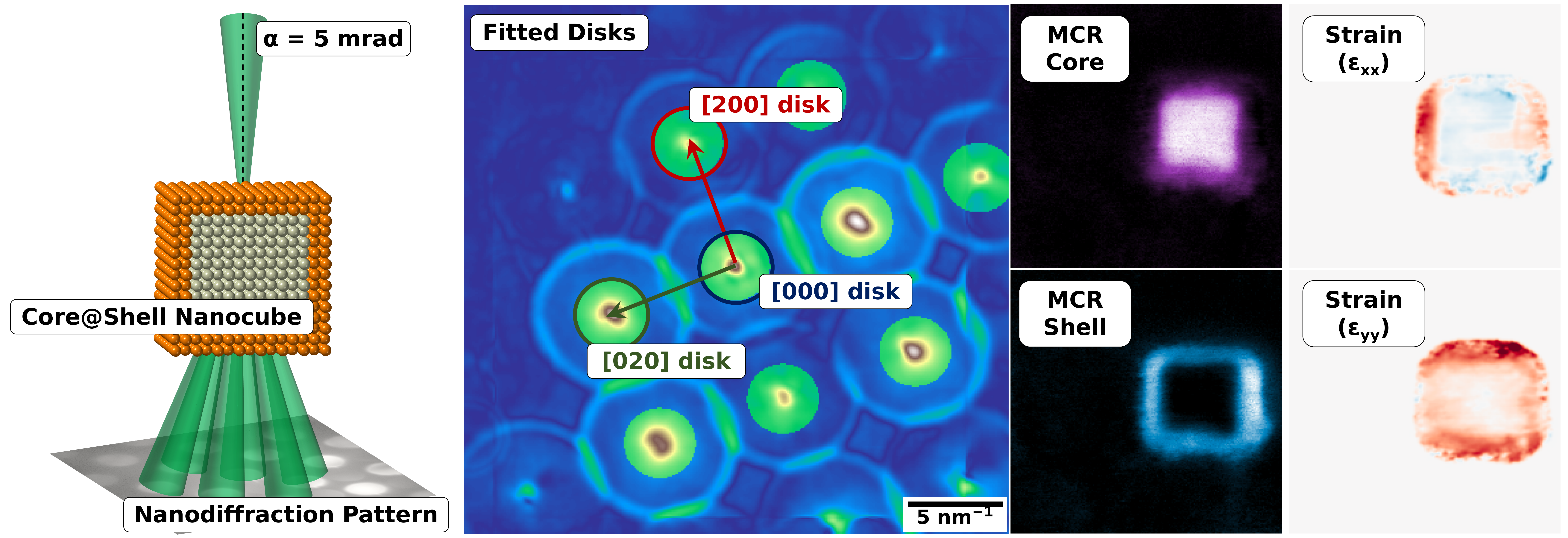}
	\end{tocentry}
\makeatletter\@input{xx.tex}\makeatother
\end{document}


\section{\label{sec:scan_angles}Effects of scan distortions on strain}
    	\begin{figure*}
    		\includegraphics[width=\textwidth]{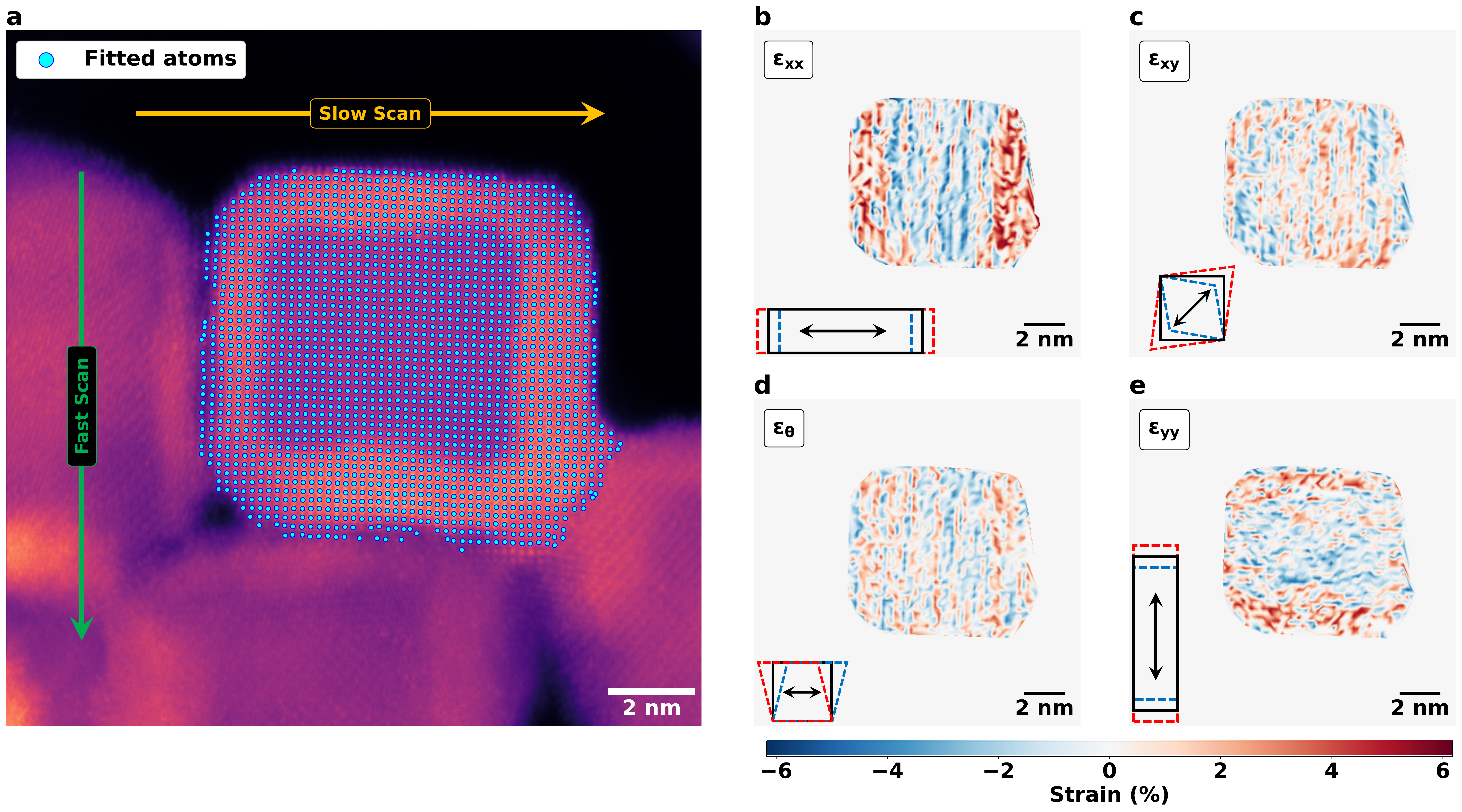}
    		\caption{\label{fig:Strain00}\textbf{Lattice strain measurement (\emph{slow scan along x-axis}). (a)} Atomic resolution image of the nanoparticle, with the fast and slow scan axes shown. The refined atom positions overlaid on \autoref{fig:Strain00}(a) as blue dots. \textbf{(b) - (e)} $\mathrm{\epsilon_{xx},\epsilon_{xy},\epsilon_{\theta}\ and\ \epsilon_{yy}}$ strain measured from the refined atom positions, with the graphically illustrated strain conventions shown in the bottom left of each individual strain map.}
    	\end{figure*}
    
        \begin{figure*}
        	\includegraphics[width=\textwidth]{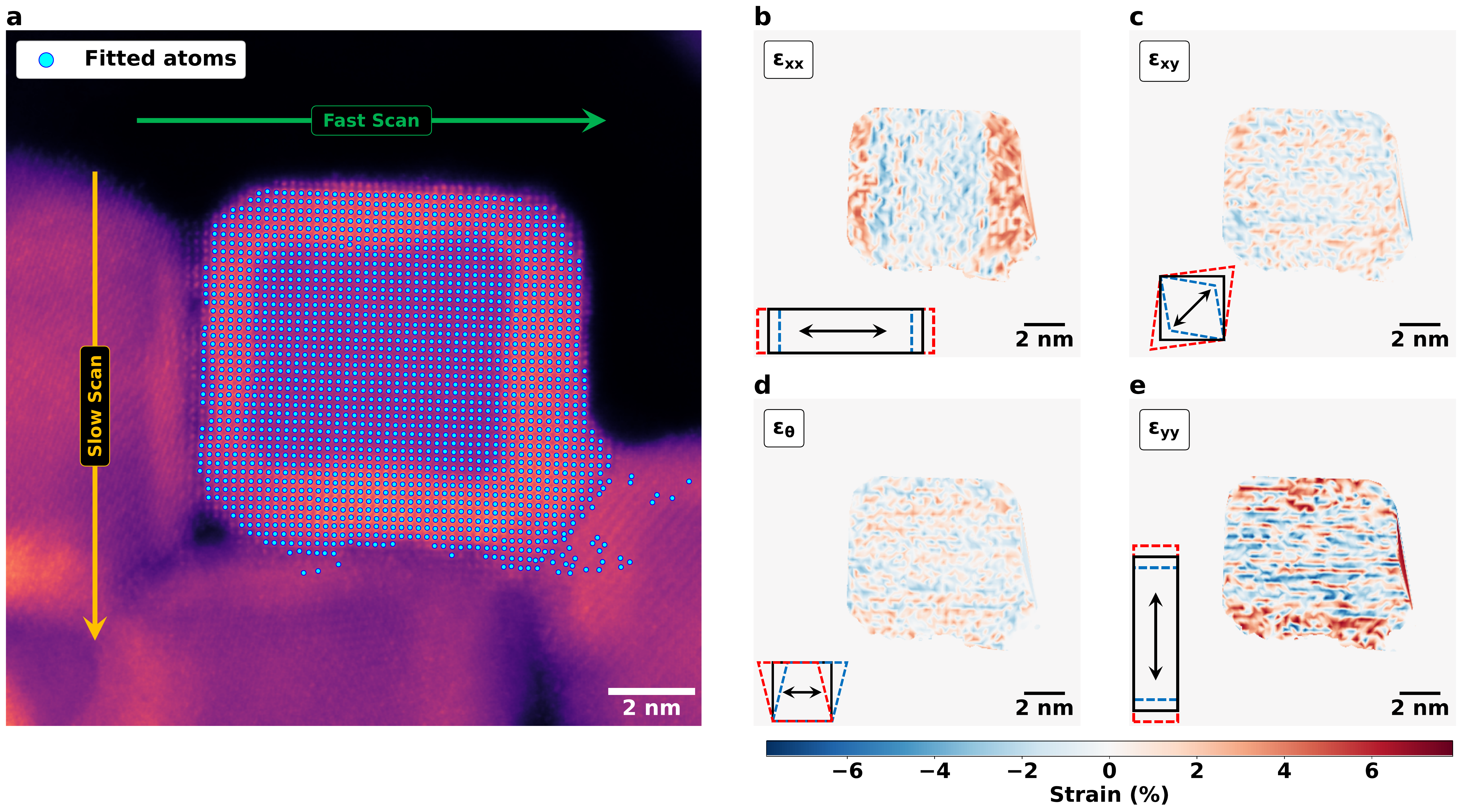}
        	\caption{\label{fig:Strain90}\textbf{Lattice strain measurement (\emph{slow scan along y-axis}). (a)} Atomic resolution image of the nanoparticle, with the fast scan and slow scan axes shown. The refined atom positions overlaid on \autoref{fig:Strain90}(a) as blue dots. \textbf{(b) - (e)} $\mathrm{\epsilon_{xx},\epsilon_{xy},\epsilon_{\theta}\ and\ \epsilon_{yy}}$ strain measured from the refined atom positions, with the graphically illustrated strain conventions shown in the bottom left of each individual strain map.}
        \end{figure*}
    
       ADF-STEM datasets were collected using two orthogonal scan directions, shown as \autoref{fig:Strain00}(a) and \autoref{fig:Strain90}(a). A flyback scanning procedure was used for both the scanning images, where the electron scans along a direction (also referred to as the \emph{fast scan direction}), and after completion of each scan line returns back to the initial scan position, shifts down by a single pixel spacing and then starts scanning the subsequent line. As could be ascertained, thus the velocity of beam movement along the direction orthogonal to the fast scan direction is almost three orders of magnitude slower.
       
       This can lead to artifacts from scan distortions, as visible for example in \autoref{fig:Strain00}(b) as stripes and striations in the $\mathrm{\epsilon_{xx}}$ strain maps. Similar confounding stripes can be observed in \autoref{fig:Strain90}(e) for the $\mathrm{\epsilon_{yy}}$ strain maps. Notably the stripes $\mathrm{\epsilon_{xx}}$  stripes in \autoref{fig:Strain00}(b) are absent in \autoref{fig:Strain90}(b) and vice-versa for the $\mathrm{\epsilon_{yy}}$ features. This indicates that rather than being strain features in the material, they arise due to scan distortions, with the stripes originating perpendicular to the slow scan directions. In order to account for such distortions, these two orthogonal scan pairs were subsequently corrected for scanning distortions using MATLAB scripts developed originally by Ophus et. al\cite{drift_corr}.
    
   \section{\label{sec:precond}Preconditioning diffraction data}
       Our two step data preconditioning routine proceeds as following:
       \begin{enumerate}
       	\item \textbf{Logarithm of diffraction pattern:} The raw diffraction pattern is flattened in intensity space by taking the logarithm of the diffraction pattern. This is because for strain mapping, we are not interested in the features inside a diffraction disk, but rather the location of the disk itself, so taking the logarithm of the data smooths out the intensity variations of the diffraction disks themselves, and decreases the intensity variations between disks. Thus, if $\mathit{C}$ is the CBED pattern, after the first step of preconditioning we obtain $\mathit{LC}$ as shown in \autoref{eq:log}
       	\begin{equation}\label{eq:log}
       	\mathit{
       	LC = \log_{10}\left( 1 + \frac{C - C_{min}}{C_{max} - C_{min}} \right)
       }
       	\end{equation}
       	As could be ascertained from \autoref{eq:log}, the pattern is normalized, so that the intensity values range from +1 to +2 to prevent taking logarithms of negative data, or values below 1. 
       	\item \textbf{Sobel-Filtering:} We subsequently Sobel filter the logarithm of the CBED pattern $\mathit{\left( LC \right)}$. The Sobel operators are two $\mathrm{3 \times 3}$ kernels used frequently for edge detection in computer vision\cite{sobel,sobel_edge}. When the kernels are convolved with an image, they give the approximate derivatives of the image along the two Cartesian directions of the image. The results from convolution with the two Sobel kernels -- $\mathit{SC_x}$ and $\mathit{SC_y}$ are given as per \autoref{eq:sobel_x} and \autoref{eq:sobel_y} respectively, where $\mathit{LC}$ is obtained as shown in \autoref{eq:log}. $\mathrm{\otimes}$ refers to convolution with a kernel.
       	\begin{equation}\label{eq:sobel_x}
       	\mathit{
       	SC_x = \begin{bmatrix}
       	-1 & 0 & 1 \\
       	-2 & 0 & 2 \\
       	-1 & 0 & 1
       	\end{bmatrix} \otimes LC
       }
       	\end{equation}
       	\begin{equation}\label{eq:sobel_y}
       	\mathit{
       	SC_y = \begin{bmatrix}
       	-1 & -2 & 1 \\
       	0 & 0 & 0 \\
       	1 & 2 & 1
       	\end{bmatrix} \otimes LC
       }                  
       	\end{equation}
       	Subsequently, we calculate the absolute magnitude of the Sobel derivative as per \autoref{eq:sobel}
       	\begin{equation}\label{eq:sobel}
       	\mathit{
       	SC = \sqrt{SC_x^2 + SC_y^2}
       }
       	\end{equation}
       \end{enumerate}
        If the first preconditioning step (\autoref{eq:log}) is not followed, then Sobel filtering will pick up disk features as intensity variations, and along with \emph{real} disk edges internal features will be highlighted too. The advantage of this two-step routine is that it is computationally relatively inexpensive to implement, but allows high-precision disk location without the need for specialized patterned condenser apertures. 
        
\begin{figure*}
        	\includegraphics[width=\textwidth]{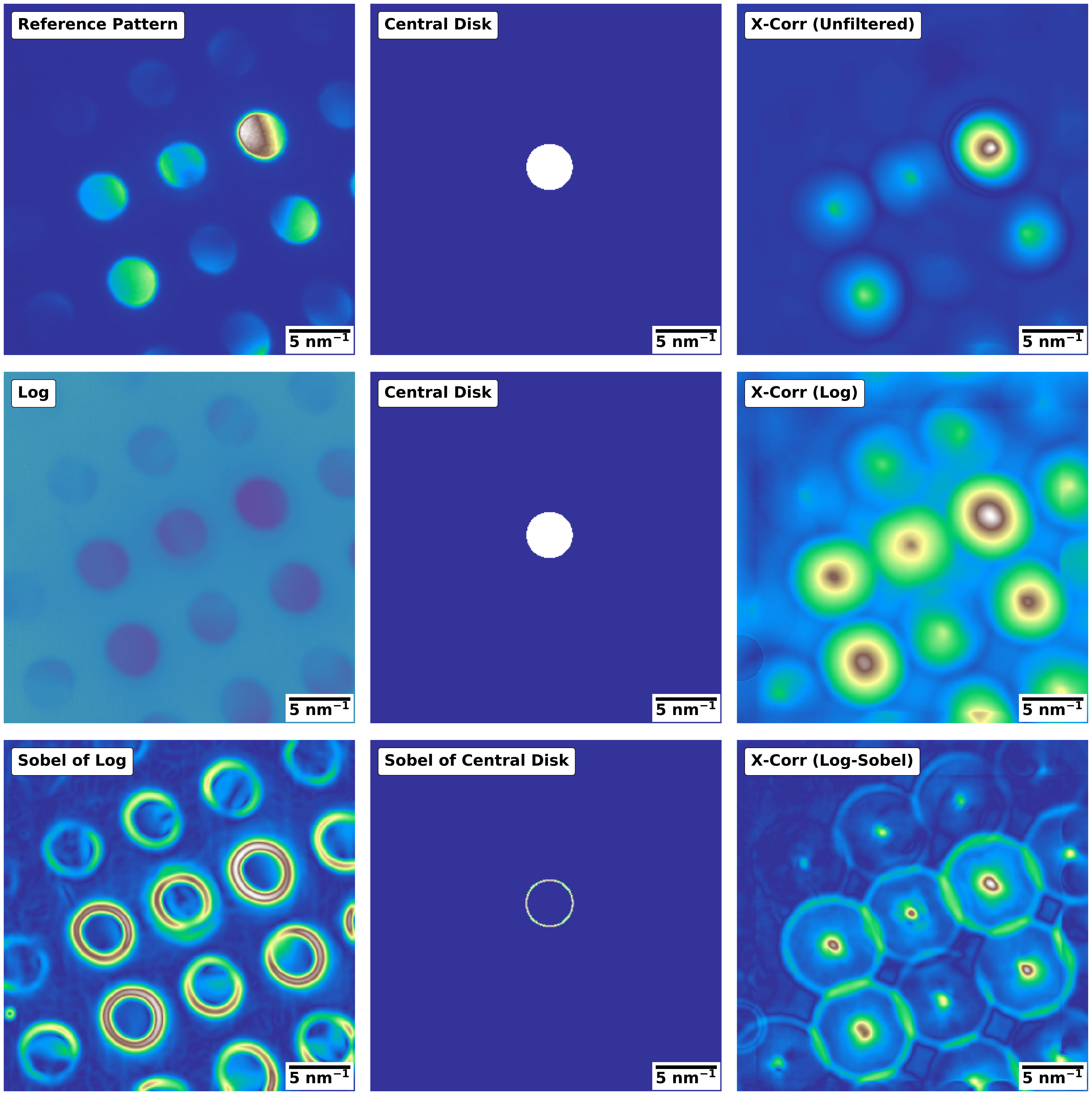}\caption{\label{fig:Filter}\textbf{Effect of preconditioning on peak sharpness}}
        \end{figure*}

        The effects of preconditioning can be visualized in \autoref{fig:Filter}, where the cross-correaltion peaks are significantly blurry for the raw datasets, sharper for logarithm of the CBED datasets and even sharper for preconditioned datasets. Additionally, as we can observe from  \autoref{fig:Filter}, preconditioning the pattern also allows for a larger number of diffraction disks to be fitted, and thus increasing the accuracy of unit cell quantification. 
        
        Similar to approaches adopted by Zeltzmann \emph{et. al.}\cite{patterned_aperture}, we followed the cross-validation (CV) approach for measuring the error of peak fitting. In this procedure the disk fitting and strain measurement is performed twice. For every dataset,apart from the central $\mathrm{\left\langle 000 \right\rangle}$, half the disks are fitted, while in the second measurement the $\mathrm{\left\langle 000 \right\rangle}$ disk and the other disks that were not fitted the first time are fitted. The calculated unit cell is compared between the two measurements - which is the CV error.
        
        We observed a CV error of raw data at \textbf{0.216\%}, for logarithm of the CBED data at \textbf{0.1962\%} and an error of \textbf{0.074\%} for preconditioned data. The preconditioned data is thus approximately 3 times more accurate than the raw data, and demonstrates performance similar to bulls-eye apertures\cite{patterned_aperture}.
     \section{\label{sec:region_roi}Region identification with MCR}  
       \begin{figure*}
       	\includegraphics[width=0.66\textwidth]{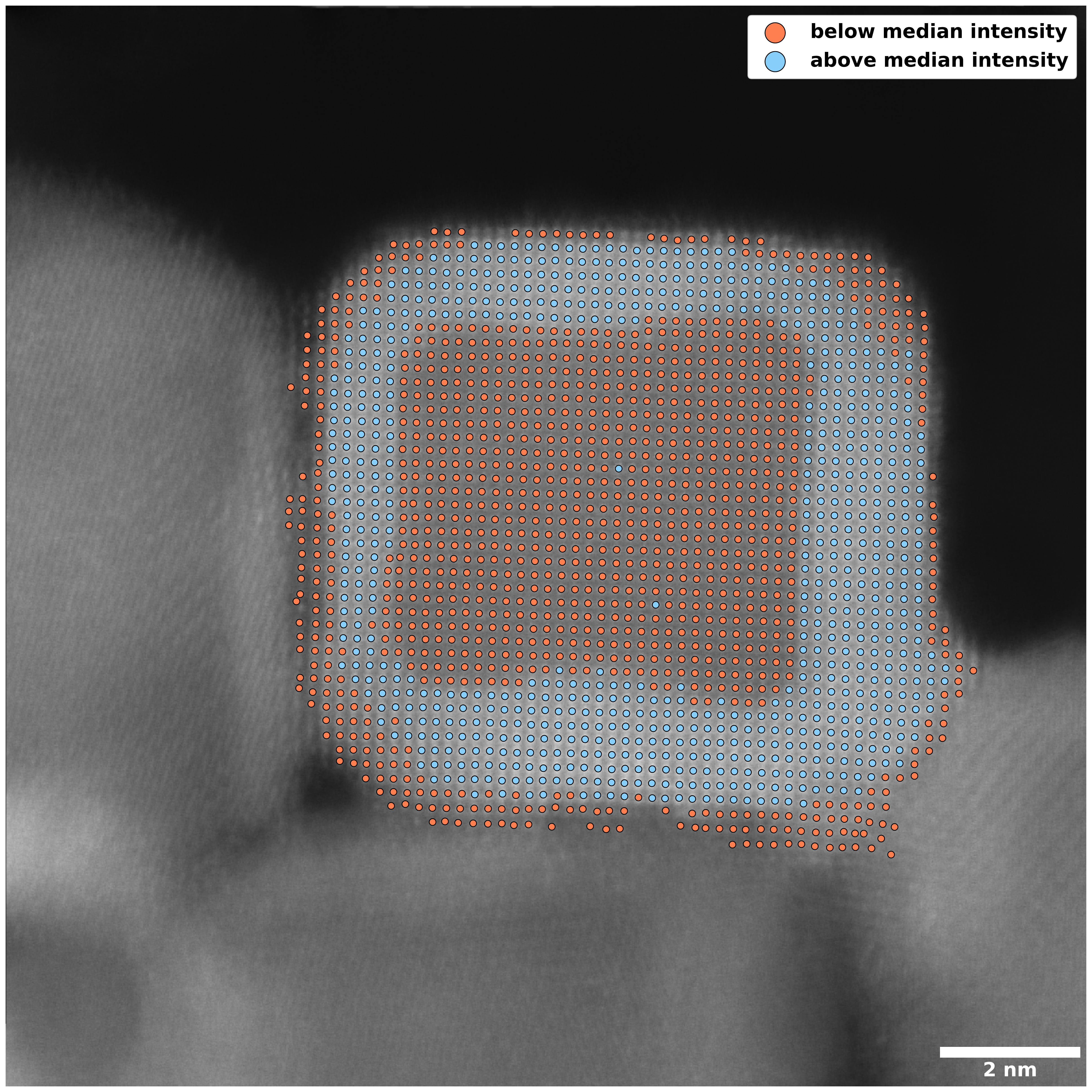}
       	\caption{\label{fig:ADF_Intensity}\textbf{Intensity distribution of atom columns in ADF-STEM}}
       \end{figure*}
       
       Due to \textit{Z contrast} in atomic resolution HAADF-STEM, distinction can be made between the core and shell of the nanoparticle through atom column intensities too, as demonstrated in \autoref{fig:ADF_Intensity}. However, the intensity of each individual atom columns depends not only on the atomic (Z)  number, but also the total number of atoms in that column. As a result, as could be observed in \autoref{fig:ADF_Intensity}, a simple partitioning of the intensities into two sets -- below and above the median intensity of the columns is close but not completely accurate.
       
       Going by this simplistic scheme, the edges of the nanocube which have lesser number of atoms in each column will be erroneously assigned to the particle core rather than the shell. However, a nanoparticle is a special case, and in many other examples like thin films the sample thickness is pretty uniform and intensity distributions can be used for region identification from ADF-STEM datasets.
       
       In 4D-STEM nanodiffraction however individual atom columns are not distinguishable so a region identification scheme would need to distinguish based on the diffraction patterns at individual scan positions itself.
              
        \begin{figure*}
        	\includegraphics[width=\textwidth]{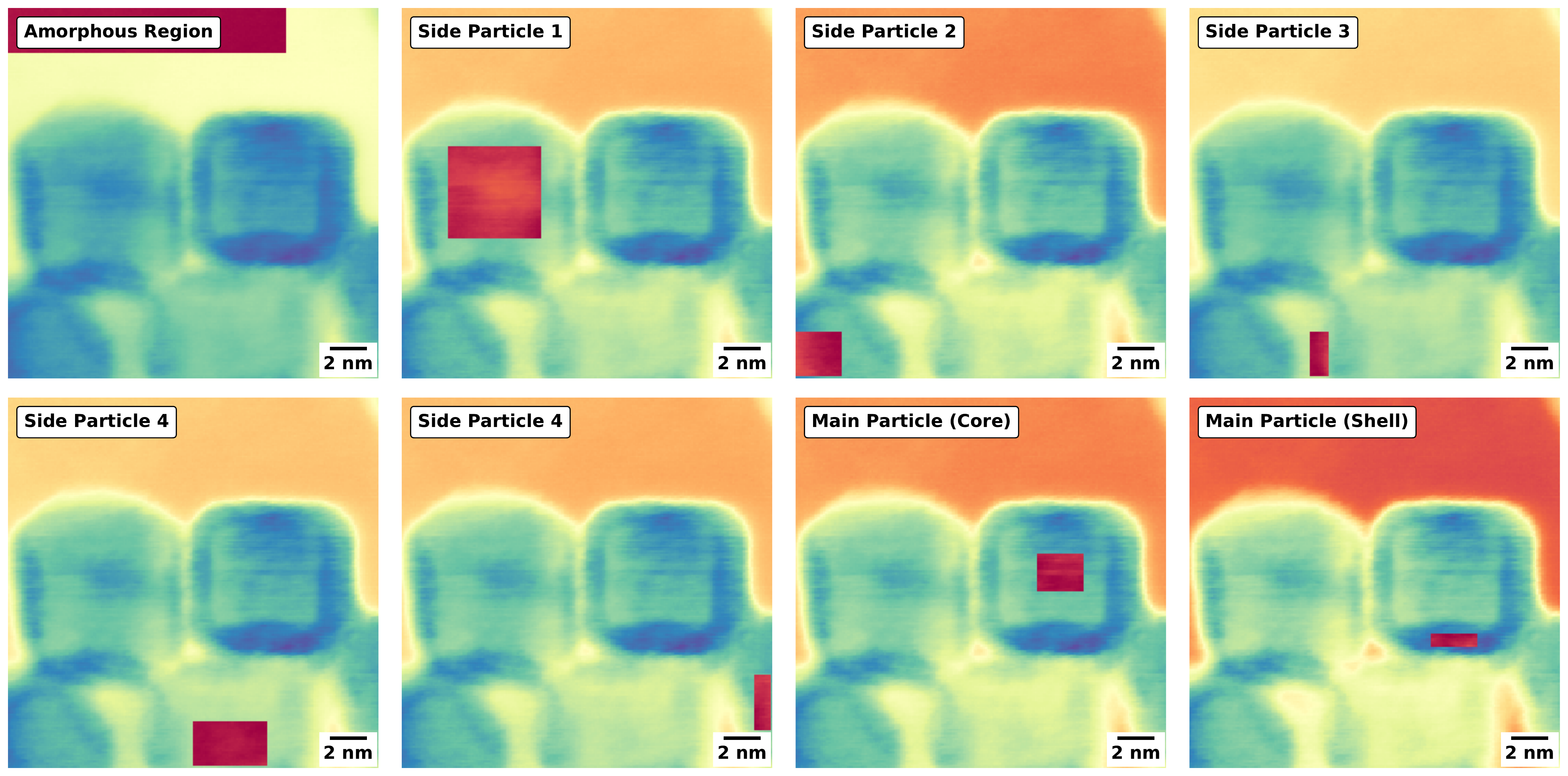}
        	\caption{\label{fig:MCR_ROI}\textbf{Regions of Interest (ROI) chosen manually for location identification with MCR, overlaid as red rectangles on the images}}
        \end{figure*}
       
       MCR requires template spectra for matching. The spectra in this case was chosen manually, by selecting a region of the sample, with the mean CBED pattern from that region being the template spectra. The regions are demonstrated in \autoref{fig:MCR_ROI}, with each neighboring particle, the amorphous region and the particle core and the particle shell being chosen as templates.
       
       Since MCR can only match 1D spectra, the CBED patterns from each region are first downsampled by a factor of 4, and then unrolled as a 1D spectra. This is then compared with the unrolled, downsampled CBED spectra from every scanning point for region identification.
        \begin{figure*}
        	\includegraphics[width=\textwidth]{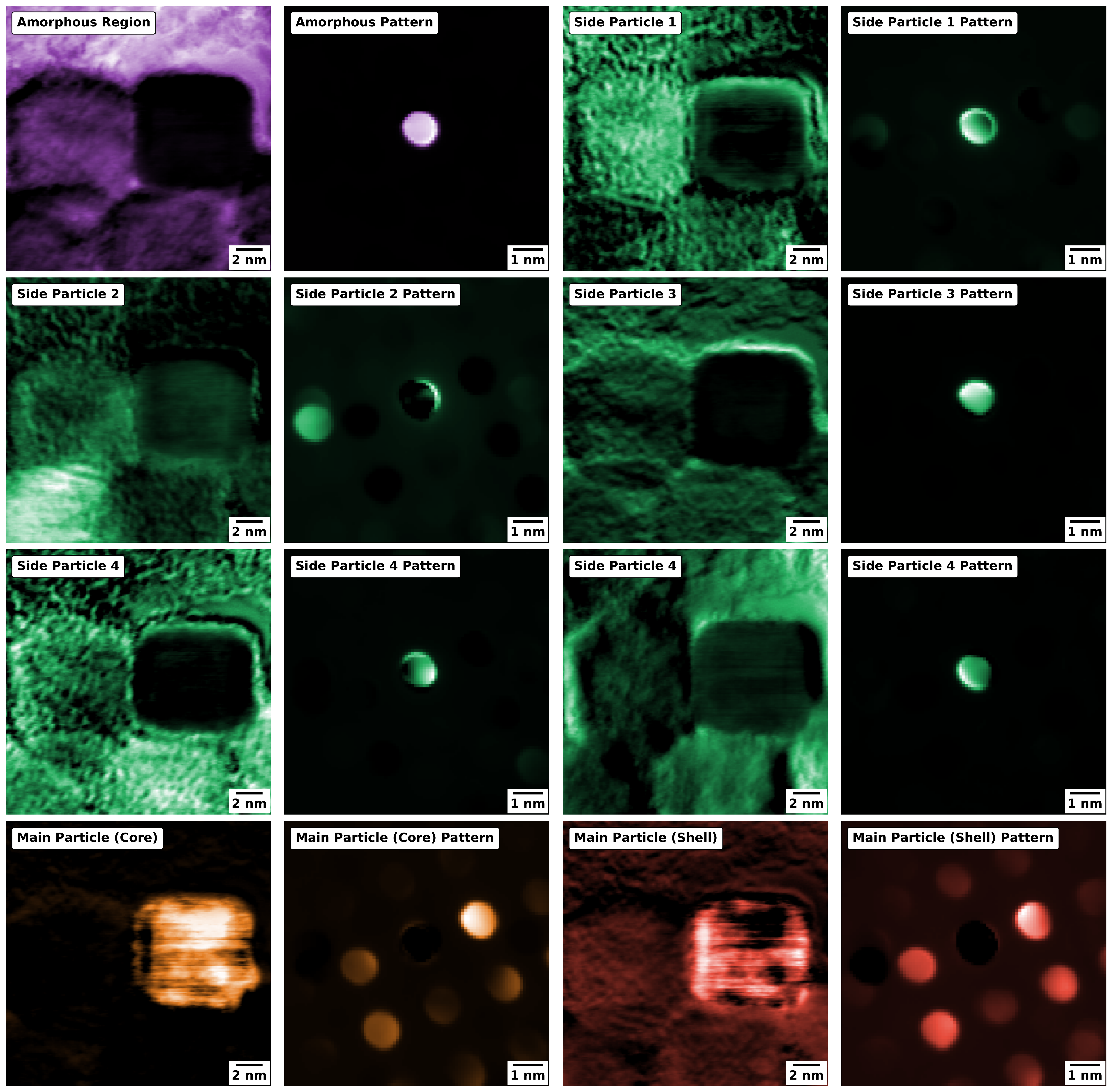}
        	\caption{\label{fig:MCR_Regions_Raw}\textbf{MCR Results on unfiltered CBED patterns}}
        \end{figure*}
    
       \begin{figure*}
	       	\includegraphics[width=\textwidth]{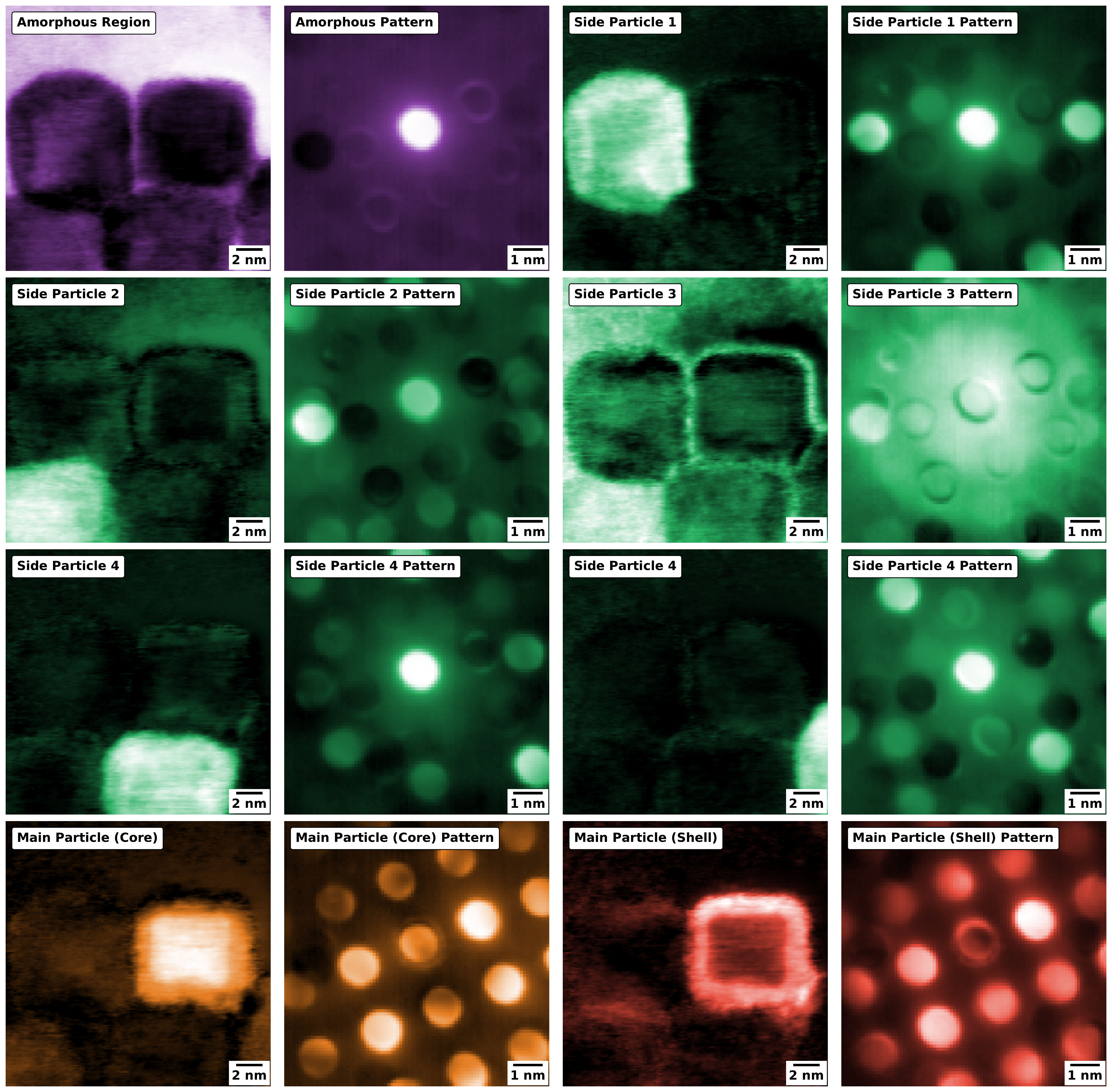}
	       	\caption{\label{fig:MCR_Regions_Log}\textbf{MCR Results on log of CBED patterns}}
       \end{figure*}
   
	    \begin{figure*}
	    	\includegraphics[width=\textwidth]{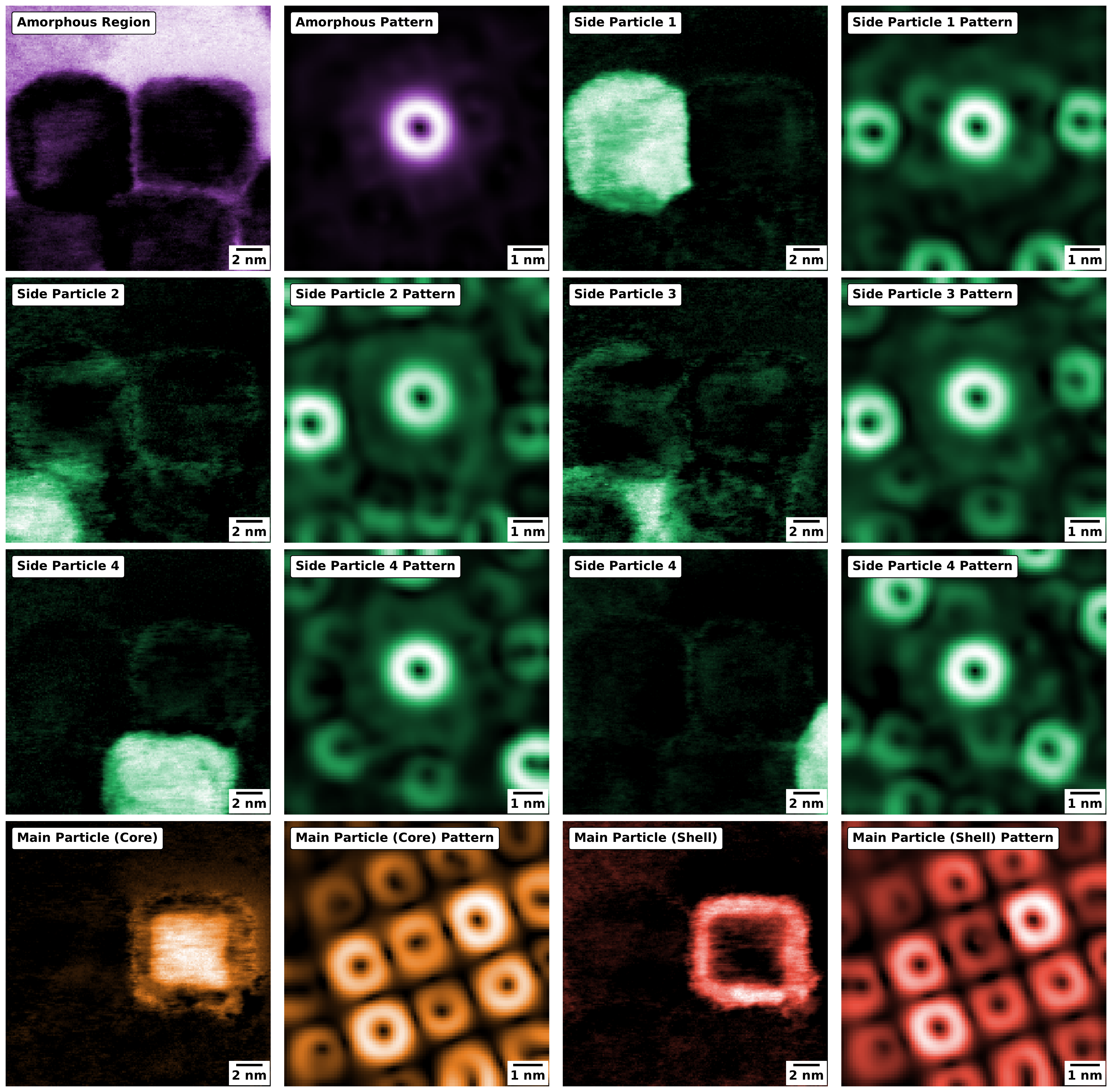}
	    	\caption{\label{fig:MCR_Regions}\textbf{MCR Results on Preconditioned CBED patterns}}
	    \end{figure*}
     
     MCR was performed on the unprocessed data (\autoref{fig:MCR_Regions_Raw}), the logarithm of the data  (\autoref{fig:MCR_Regions_Log}) and the preconditioned data  (\autoref{fig:MCR_Regions}). Similar to the advantages of data preconditioning for strain mapping, we observed MCR actually performed better on logarithm of CBED patterns rather than the raw patterns, with preconditioned data outperforming both of them for region identification. 
    
   	\bibliography{manuscriptReview}
\makeatletter\@input{yy.tex}\makeatother